\documentclass[aps,prl,reprint,footnotebib,amsfonts,amsmath,amssymb,superscriptaddress]{revtex4-2}

\usepackage{graphicx}
\usepackage{color}
\usepackage{tabularx} 

\usepackage[bookmarks=true,%
    colorlinks=true,%
    linkcolor=blue,%
    citecolor=blue,%
    filecolor=blue,%
    menucolor=blue,%
    urlcolor=blue,%
    pdfduplex=DuplexFlipLongEdge,%
    breaklinks=true]{hyperref}

\usepackage{lineno}

\usepackage{latexsym,exscale,enumerate,comment, bbold}
\usepackage{amsthm,amscd}
\usepackage{subfigure}
\usepackage{slashed}
\usepackage{braket}
\usepackage[all,knot]{xy}
\usepackage{tikz}
\usepackage{upgreek}
\usepackage{bm}

\def\bea{\begin{eqnarray}}
\def\eea{\end{eqnarray}}
\def\nn{\nonumber}
\def\ba{\begin{array}}
\def\ea{\end{array}}
\def\nn{\nonumber}
\def\Tr{\text{Tr}}

\newcommand{\eq}[2]{
	\begin{equation}
	#1 \label{#2}
	\end{equation}
}

\usepackage[normalem]{ulem}

\makeatletter

\begin{document}
\title{Quantum Charge-4e Superconductivity and Possible Deconfined Pseudocriticality in the Attractive SU(4) Hubbard Model}
\author{Zhou-Quan Wan}
\thanks{These two authors contributed equally to this work.}
\affiliation{Center for Computational Quantum Physics, Flatiron Institute, New York, NY 10010, USA}

\author{Huan Jiang}
\thanks{These two authors contributed equally to this work.}
\affiliation{Department of Physics and Engineering Physics, Tulane University, New Orleans, Louisiana 70118, USA}

\author{Xuan Zou}
\affiliation{Chen-Ning Yang Institute for Advanced Study (YIAS), Tsinghua University, Beijing 100084, China}

\author{Shiwei Zhang}
\affiliation{Center for Computational Quantum Physics, Flatiron Institute, New York, NY 10010, USA}

\author{Shao-Kai Jian}
\email{sjian@tulane.edu}
\affiliation{Department of Physics and Engineering Physics, Tulane University, New Orleans, Louisiana 70118, USA}

\maketitle 
\textbf{Unlike conventional charge-2e superconductors, a charge-4e superconductor exhibits long-range coherence of electron quartets rather than Cooper pairs.
Clear zero-temperature realizations of charge-4e superconductivity remain rare.
Here, we investigate the zero-temperature phase diagram of the attractive SU(4) Hubbard model with numerically exact, large-scale quantum Monte Carlo (QMC) simulations overcoming major technical hurdles.
We identify both charge-2e and charge-4e superconducting phases.
Upon increasing interaction, charge-2e long-range order vanishes, while charge-4e correlations remain robust and converge with system size, establishing a quartet-condensed phase.
Interestingly, across the charge-2e--charge-4e transition, single electrons remain gapped, while charge-2e correlations exhibit an effective anomalous dimension substantially exceeding the conventional O(6) Landau theory prediction.
These features motivate a fractionalized framework in which the physical charge-2e order parameter is a composite field coupled to an emergent non-Abelian gauge structure.
We formulate an Sp(4) gauge-Higgs theory associating the charge-2e and charge-4e states with Higgs and confined phases, respectively.
A one-loop analysis of an SU($N_f$) generalization exhibits a fixed-point collision, motivating deconfined pseudocriticality as a possible interpretation of the observed scaling.
Our results establish charge-4e superconductivity as a bona fide zero-temperature phase, provide a simple model for future studies in a numerically exact framework, and suggest an unconventional route to superconducting criticality.
}

\vspace{-0.5em}
\section{Introduction}
\vspace{-0.5em}
Superconductivity (SC) is the canonical example of emergent macroscopic coherence, yet it is usually understood in terms of the condensation of charge-$2e$ Cooper pairs.
Charge-$4e$ SC generalizes this notion: long-range phase coherence is carried by electron quartets, implying a fundamental flux quantum $hc/4e$ and qualitatively new vortices and Josephson responses compared to conventional superconductors \cite{schlottmann1994ground,babaev2004superconductor,wu2005competing,capponi2008molecular,berg2009charge}. 
The most well-developed route to charge-$4e$ order is as a vestigial descendant of more conventional pairing, such as from the partial melting of a pair-density-wave superconductor, where fluctuating finite-momentum pair order can destroy charge-$2e$ coherence while preserving quartet coherence \cite{berg2009charge,Radzihovsky2011}. 
More broadly, higher-charge condensates have been proposed in multicomponent superconductors and superfluids, including nematic and chiral settings \cite{herland2010phase,fernandes2021charge,jian2021charge,hecker2023cascade,hecker2024local,volovik2024,zou2025emergence}. 
Such states are also predicted to emerge in twisted and kagome materials~\cite{zhou2022chern,maccari2023prediction,liu2023charge,wu2024dwave}. 
Experimental motivation has intensified with recent reports of multi-charge flux quantization in kagome-superconductor ring devices~\cite{ge2024charge}.

Despite rapid progress, charge-$4e$ SC has most often been discussed as a finite-temperature vestigial order of nearby charge-$2e$ pairing.
While a growing set of microscopic examples demonstrates that charge-$4e$ SC can appear as a zero-temperature phase stabilized by genuine quartet condensation~\cite{Lecheminant2005,liu2023quartet,soldini2024charge,gao2026primary}, microscopic realizations of zero-temperature charge-$4e$ ground states remain relatively few and a unifying picture is still developing~\cite{jiang2017charge,you2018symmetric,khalaf2022symmetry,gnezdilov2022solvable,gao2025,samoilenka2026microscopic}. 
At a more fundamental level, the quantum phase transitions into and out of quartet-condensed ground states remain largely uncharted. 

\begin{figure}
    \centering
    \includegraphics[width=1.0\linewidth]{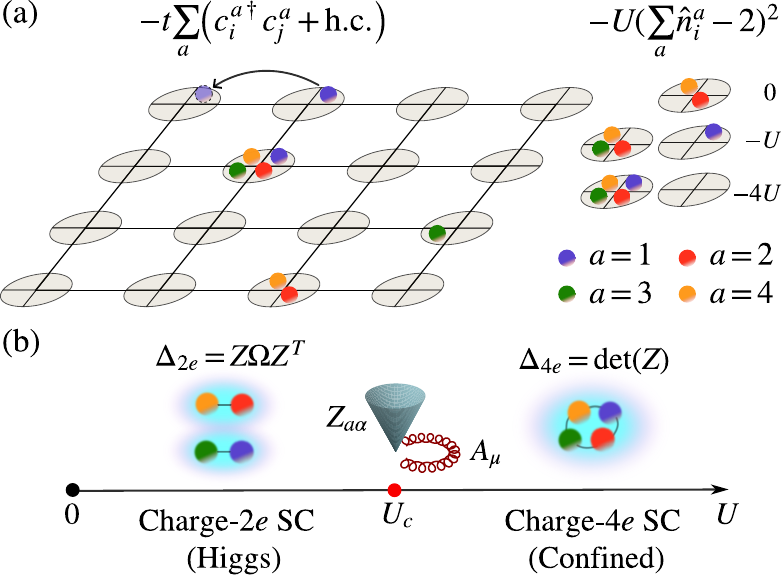}
    \caption{\textbf{Model and zero-temperature phase diagram.} (a) Schematic of the attractive SU(4) Hubbard model on the square lattice, and the interaction energies of all configurations on a site. 
    (b) Zero-temperature phase diagram as a function of $U$, showing a transition from charge-$2e$ to charge-$4e$ superconductivity as \(U\) increases. In the proposed Sp(4) gauge-Higgs description, these states correspond to Higgs and confined phases, respectively. Deconfined pseudocriticality provides a possible interpretation of the transition.}
    \label{fig:schematic}
\end{figure}

Here we address these issues in the attractive SU(4) Hubbard model, using numerically exact determinant quantum Monte Carlo (DQMC) simulations. 
Despite the absence of a sign problem with DQMC in this model, reliable computations remain challenging due to technical hurdles, most notably an infinite-variance problem~\cite{shihaoPRE2016,wan2025bridge} in evaluating charge-$4e$ correlations. 
Overcoming this issue~\cite{wan2025bridge}, we perform large-scale simulations (over $2700$ lattice sites with more than $1300$ fermions) to map out the zero-temperature phase diagram of the SU(4) Hubbard model.
Our simulations reveal both a conventional charge-$2e$ SC phase and a distinct regime with robust charge-$4e$ SC correlations, separated by a sharp interaction-tuned transition, as illustrated in Fig.~\ref{fig:schematic}.

We next examine the unconventional quantum critical behavior between the charge-$2e$ and charge-$4e$ superconducting regimes. 
Single-electron excitations remain gapped across the transition, showing that it is not driven by the re-emergence of a Fermi surface but by a reorganization of collective modes. 
In addition, the finite-size scaling of the charge-$2e$ correlations exhibits pronounced systematic drift, and the effective exponents extracted from size-pair analyses remain strongly size dependent.
On symmetry grounds, one might anticipate a conventional Landau--Ginzburg--Wilson (LGW) description; however, the anomalous dimension predicted by such a theory is far smaller than that implied by our data. 
This mismatch disfavors a conventional LGW account and motivates a fractionalized description in which the physical charge-$2e$ order parameter is a composite operator. 
We therefore formulate a non-Abelian Sp(4) gauge-Higgs theory for a matrix Higgs field that carries global SU($N_f$) quantum numbers. 
In such theories, deconfined critical points can arise beyond the Landau paradigm \cite{fradkin1979phase,hikami1980non,senthil2004deconfined}, and renormalization-group (RG) flows can exhibit fixed-point collisions and pseudocritical behavior \cite{wang2017deconfined,gorbenko2018walking,bonati2024charged,ihrig2019abelian,RCMa2020,Nahum2020,Zhu2023,zou2025unraveling}. 
We show that this mechanism provides an organizing framework for the observed finite-size drift and yields collision-point exponents for comparison with our numerics.
This framework suggests a possible route to quantum criticality between charge-$2e$ and charge-$4e$ superconductors, in which fractionalization and an emergent non-Abelian gauge structure can give rise to deconfined pseudocritical behavior.

\vspace{-0.5em}
\section{Results}
\vspace{-0.5em}
\noindent{\bf Phase Diagram and Quantum Charge-4e SC}\\
The SU(4) Hubbard model on a square lattice reads
\begin{equation}
\hat H= -t \sum_{a,\langle ij\rangle}({c_i^a}^\dagger c_j^a + \text{h.c.})  - U \sum_{i}(\sum_{a}\hat n_i^a-2 )^2 \,,\label{eq:hamiltonian}
\end{equation}
where ${c_i^a}^\dagger$ ($c_i^a$) denotes the electron creation (annihilation) operator at site $i$ with the flavor index $a =1,2,3,4$ and $\hat n_i^a = {c_i^a}^\dagger c_i^a$ is the electron density operator. 
Throughout this work, we set $t = 1$ as the unit of energy.
As illustrated in Fig.~\ref{fig:schematic}(a), $t$ and $U$ are the nearest-neighbor (NN) hopping amplitude and the Hubbard interaction strength, respectively. 
Using large-scale DQMC simulations on lattices up to $52\times 52$, we study the model at 1/8 filling ($L^2/2$ electrons), with methodological details provided in Methods.

\begin{figure}
    \centering
    \includegraphics[width=1.0\linewidth]{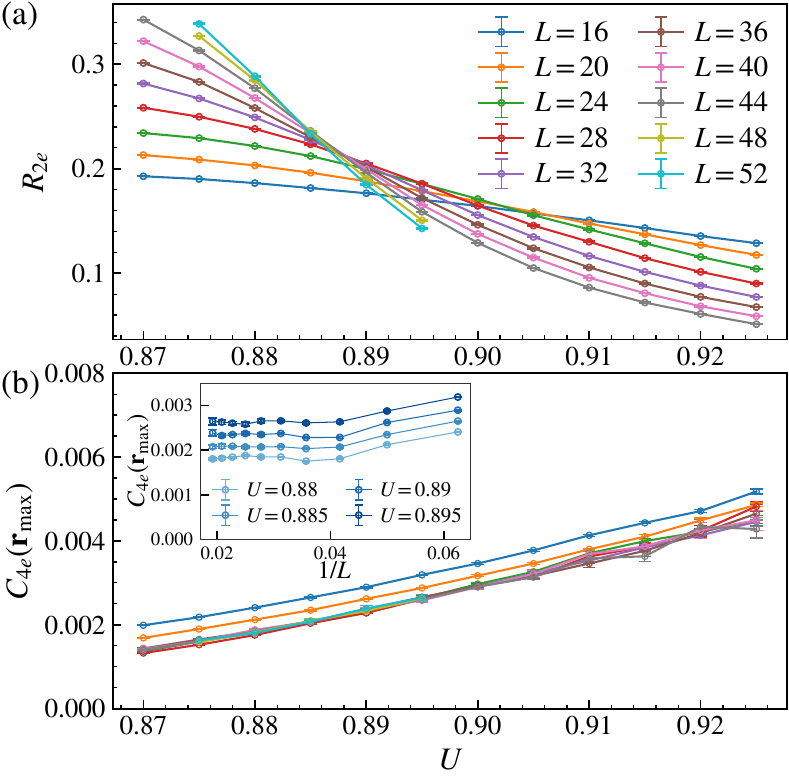}
    \caption{\textbf{Transition from charge-$2e$ to charge-$4e$ superconductivity.} (a) Correlation ratio of the charge-$2e$ order parameter, $R_{2e}$, as a function of interaction strength $U$ for different system sizes $L$. 
    The crossing behavior signals the suppression of charge-$2e$ SC order beyond a critical coupling $U_c$.
    (b) Charge-$4e$ correlation $C_{4e}(\mathbf{r}_{\text{max}})$ at the largest lattice separation $\mathbf{r}_{\text{max}}=(L/2,L/2)$, as a function of $U$.  
    In contrast to the charge-$2e$ order, the charge-$4e$ correlations remain robust into the strong-coupling side, confirming the persistence of charge-$4e$ SC long-range order. 
    Inset: $C_{4e}(\mathbf{r}_{\text{max}})$ as a function of $1/L$ for representative couplings across the critical region. 
    The data exhibit clear convergence as $1/L\rightarrow 0$, indicating finite charge-$4e$ correlations in the thermodynamic limit.}   \label{fig:numeric1}
\end{figure}

To probe charge-$2e$ SC, we define the charge-$2e$ order operator $\Delta_{2e}(i) =  c_i^a c_i^b$ (with $a\neq b$, and other channels related by SU(4) symmetry), and compute the corresponding structure factor
\begin{equation}
    S_{2e}(\mathbf{k})=\frac{1}{N} \sum_{i,j} e^{i\mathbf{k}\cdot(\mathbf{r}_i-\mathbf{r}_j)} \left\langle \Delta_{2e}(i)\Delta_{2e}^\dagger(j)+\mathrm{h.c.}\right\rangle\,.\label{eq:structure_factor}
\end{equation}
We further define the correlation ratio $R_{2e}=1-\frac{S_{2e}(\mathbf{k}_{\rm min})}{S_{2e}(\mathbf{0})}$, where $\mathbf{k}_{\rm min}=(2\pi/L,0)$ is the smallest nonzero momentum.
As shown in Fig.~\ref{fig:numeric1}(a), $R_{2e}$ exhibits a clear crossing as $U$ increases, indicating that the charge-$2e$ order becomes disordered for $U>U_c$.

We next consider the charge-$4e$ order parameter, defined as $\Delta_{4e}(i)= c_i^1c_i^2c_i^3c_i^4$, and evaluate its correlation at the maximum separation $\mathbf{r}_{\rm max}=(L/2,L/2)$, $C_{4e}(\mathbf{r}_{\rm max})=\frac{1}{N}\sum_i\langle \Delta_{4e}(i+\mathbf{r}_{\rm max})\Delta_{4e}^\dagger(i)+\mathrm{h.c.}\rangle$. 
A severe infinite-variance problem is present in this model, and it is essential to employ the exact bridge link method to obtain reliable estimates~\cite{shihaoPRE2016, wan2025bridge}. 
As shown in Fig.~\ref{fig:numeric1}(b), $C_{4e}(\mathbf{r}_{\rm max})$ converges with system size. Additional data on $C_{4e}(\mathbf{r})$ are illustrated in Methods. 
The inset shows that $C_{4e}(\mathbf{r}_{\rm max})$ approaches a finite value as $1/L\rightarrow 0$. 
Moreover, its magnitude increases monotonically with $U$ in this regime, indicating robust long-range charge-$4e$ correlations even after the charge-$2e$ order is destroyed.
These results establish a quantum phase transition from a charge-$2e$ to a charge-$4e$ SC phase at strong coupling.
Complementary momentum-resolved measurements show no detectable finite-momentum pairing or competing charge-density-wave (CDW) order on the charge-$4e$ side up to at least $U=0.925$, consistent with the dilute-limit strong-coupling picture (see Supplementary Information for the strong-coupling analysis and density results at larger $U$).

\begin{figure}
    \centering
    \includegraphics[width=1.0\linewidth]{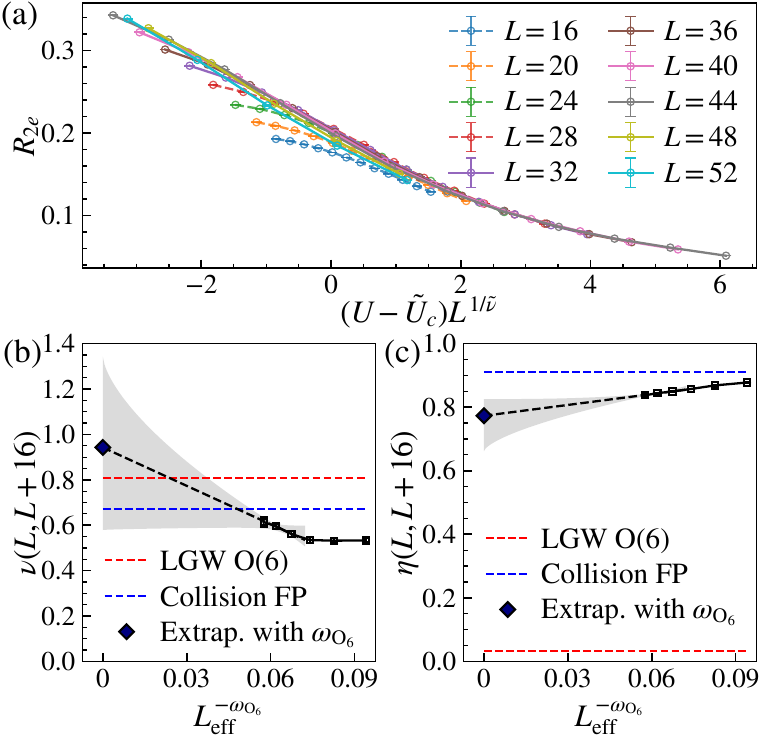}
    \caption{\textbf{Finite-size scaling and effective critical exponents.} (a) Finite-size scaling collapse of the charge-$2e$ correlation ratio $R_{2e}$, plotted as a function of $(U-\tilde{U}_c)L^{1/\tilde{\nu}}$, using parameters $\tilde{U}_c$ and $\tilde{\nu}$ estimated using data with $L\geq 32$. While the data exhibit approximate collapse, systematic deviations between different system sizes remain visible, indicating the absence of a consistent single-parameter scaling description.
    (b,c) Effective critical exponents extracted from size pairs $(L, L+16)$, with $L_{\mathrm{eff}}=\sqrt{L(L+16)}$: (b) correlation-length exponent $\nu(L,L+16)$ and (c) anomalous dimension $\eta(L,L+16)$. 
    Both quantities display pronounced size dependence. Diamonds show representative extrapolations using the O(6) correction exponent $\omega_{O_6}=0.758$~\cite{Hasenbusch2022prb}, while gray bands indicate sensitivity to the extrapolation scheme (see SI).
    Horizontal dashed lines indicate critical exponents of the LGW O(6) theory (red) and generalized one-loop critical exponents at the collision fixed point of the Sp(4) gauge-Higgs model: $\eta^\ast\approx0.91$ and $\nu^\ast\approx0.67$ (blue).} \label{fig:numeric2}
\end{figure}

\vspace{10pt}
\noindent{\bf Anomalous Criticality at the $4e$--$2e$ Transition}\\
Now we turn to the critical behavior at the $4e$--$2e$ transition. 
In Fig.~\ref{fig:numeric2}(a), we present a finite-size scaling collapse of the correlation ratio $R_\text{2e}$. 
While an approximate collapse can be achieved, it is not stable: even when restricting to larger system sizes, noticeable deviations between different system sizes persist.
Therefore, instead of relying on data collapse, we extract effective critical exponents from size-pair crossing analysis.

Specifically, for a pair of system sizes $(L_1, L_2)$, the effective correlation-length exponent $\nu(L_1,L_2)$ is obtained from the scaling of the slope $\partial_U R_{2e}$ at the crossing point $U_c(L_1,L_2)$, while the effective anomalous dimension $\eta(L_1,L_2)$ is extracted from the scaling of $m_{2e}$ evaluated at the same point, with $m_{2e}\equiv S_{2e}(\mathbf{0})/L^2$. 
Details are provided in the Supplementary Information (SI).
In Fig.~\ref{fig:numeric2}(b,c), we find that $\nu(L,L+16)$ and  $\eta(L,L+16)$ exhibit a systematic drift with increasing system size. 
These results indicate strong finite-size effects, preventing a definitive distinction between a continuous transition with strong corrections to scaling and a weakly first-order transition.
Nevertheless, the effective exponents still provide useful quantitative characterization of the scaling behavior within accessible system sizes.
We therefore examine extrapolation sensitivity using
$\Delta L=8,12,16,20$ and various correction powers
(see SI). 
The crossing extrapolations locate the transition near $U_c\simeq0.878$, while the scanned exponent ranges are $\nu_\infty\in[0.579,1.342]$ and $\eta_\infty\in[0.661,0.825]$.
We emphasize that these should be viewed as effective thermodynamic extrapolations of the accessible finite-size data, rather than definitive asymptotic critical exponents, since pseudocriticality can produce long preasymptotic scaling windows.

To gain further insight into the nature of this quantum phase transition, we next turn to a theoretical analysis.
In the weak coupling limit, the electron Fermi surface is unstable against the attractive Hubbard interaction and forms conventional charge-$2e$ SC, as observed in the DQMC simulation.
In the opposite strong-coupling limit, the large attractive Hubbard interaction projects onto the two lowest-energy states, namely the empty site and the fully occupied quartet, as shown in Fig.~\ref{fig:schematic}(a).
Upon turning on fermion hopping within this low-energy manifold, a strong-coupling expansion yields an effective Bose-Hubbard model, in which the hard-core boson represents an electron quartet (see SI). 
In this language, the bosonic superfluid directly corresponds to charge-$4e$ SC in the original fermionic model, consistent with numerical results. 
In addition, the DQMC simulation shows that the single electron remains gapped throughout the transition, with details given in the SI.

Taken together, the weak- and strong-coupling analyses show that single electrons are not the natural low-energy degrees of freedom. 
Hence, guided by symmetry, we formulate a bosonic LGW description of a direct transition between charge-$4e$ and charge-$2e$ SC phases. 
Although the global U(1) symmetry is broken in both phases, the charge-$4e$ state retains an internal SU(4) symmetry because the condensed electron quartet is an SU(4) singlet. 
By contrast, the charge-$2e$ superconductor develops a pair condensate that spontaneously breaks SU(4) down to Sp(4), consistent with our numerics.
Noting the isomorphism SU(4)/Sp(4) $\cong$ SO(6)/SO(5), we conclude that, within a conventional LGW scenario, the critical theory would be captured by an SO(6) non-linear sigma model. 
Supporting numerical results and the derivation of the LGW description are provided in the SI.
Yet the order-parameter anomalous dimension from the SO(6) universality class is far smaller than what we observe at the $4e$--$2e$ transition. 
Monte Carlo simulations~\cite{loison1999a} and $4-\epsilon$ RG analyses to six loops~\cite{antonenko1995critical} both yield $\eta_{\rm O(6)} \approx 0.031$, which is substantially smaller than the effective anomalous dimension found in Fig.~\ref{fig:numeric2}(c). 
This pronounced discrepancy in $\eta$ therefore disfavors an LGW O(6) description of the $4e$--$2e$ transition.

\vspace{10pt}
\noindent{\bf Fractionalization and Sp(4) Gauge Theory}\\
The discrepancy with a conventional LGW description strongly motivates a deconfined quantum critical scenario. 
In particular, the unusually large effective anomalous dimension suggests that  the charge-$2e$ order parameter may be a composite operator built from more elementary degrees of freedom. 
A natural fractionalized formulation introduces a matrix field $Z_{a\alpha}$, which transforms as a fundamental under the global SU(4) symmetry (index $a$) and simultaneously as an antifundamental under an emergent local Sp(4) gauge structure (index $\alpha$),
\begin{eqnarray}
    Z \rightarrow U Z V^\dagger\,, \quad U \in {\rm SU}(4)\,, \quad V \in {\rm Sp}(4)\,.
\end{eqnarray}

When $Z$ condenses as $Z_{a\alpha} \propto \delta_{a\alpha}$, it locks the global SU(4) index to the gauge Sp(4) index, thereby breaking the global SU(4) symmetry down to Sp(4). 
In addition to its SU(4) quantum numbers, $Z$ carries a global U(1) charge.  
Using the Sp(4)-invariant antisymmetric tensor, 
$\Omega = \left(\begin{array}{cc}
        0 & I_{2} \\
        - I_2 & 0
    \end{array}\right)$, 
with $I_2$ denoting the $2\times 2$ identity matrix, the gauge-invariant charge-$2e$ order parameter is $\Delta_{2e} = Z \Omega Z^T$. 
By contrast, the gauge-invariant charge-$4e$ operator is $\Delta_{4e} = \det Z$, which is an SU(4) singlet. 
It can therefore remain finite on both sides of the transition, consistent with our numerical observation that the charge-$4e$ order parameter stays nonzero across the transition.

With the fractionalized field $Z$, we propose a non-Abelian Sp(4) gauge-Higgs description at long wavelengths: the charge-$2e$ superconductor is the Higgs phase with the required symmetry breaking, whereas a gapped $Z$ leaves a confining Sp(4) gauge sector, naturally identifying the charge-$4e$ state with the confined phase. 
The $4e$--$2e$ transition is therefore a natural candidate for deconfined quantum criticality, described by
\begin{align}
     \mathcal L_{\rm crit} &= \Tr(|D_\mu Z|^2) + r \Tr(Z^\dagger Z) + \frac1{16}  \Tr (F^{\mu\nu} F_{\mu\nu})\nonumber \\
    & + u_1 [\Tr (Z^\dagger Z)]^2 + u_2 \Tr[(Z^\dag Z)^2]  +  v  \Tr[ |Z \Omega Z^T |^2 ] \nonumber \\
    & + \lambda (\det Z + \text{h.c.})\,, 
\end{align}
where $D_\mu Z= \partial_\mu Z - i g Z A_\mu$ is the covariant derivative and  $F_{\mu\nu} = \tfrac{i}{g}[D_\mu, D_\nu] $ is the corresponding field strength. 
The gauge field $A_\mu$ takes values in the Lie algebra of Sp(4). 
The parameters $u_1$, $u_2$, $v$, and $\lambda$ denote the coupling strengths of the quartic interactions. 
Here the determinant term is written in the background of the charge-$4e$ condensate with a chosen phase.

It is well established that non-Abelian gauge-Higgs theories can enter a deconfined regime when the number of Higgs flavors exceeds a critical value~\cite{hikami1980non,bonati2024charged}.
We thus promote the $Z$ field to transform as a fundamental of a global SU($N_f$) symmetry and analyze the resulting theory in a $4-\epsilon$ expansion. 
In the SU($N_f$) continuation, we retain the quartic sector parametrized by $(u_1,u_2,v)$.
The four-flavor determinant is absent from this sector for $N_f>4$ by SU($N_f$) symmetry~\footnote{
The determinant interaction is retained in the physical action, Eq.~(4). 
The generalized RG analysis examines the fixed-point collision in the common quartic sector; it does not determine the flow of the additional determinant coupling at $N_f=4$.}.
We then obtain the RG equations for $\bm w = (g^2, u_1, u_2 , v) $ as $\frac{{\rm d} \bm w}{{\rm d}l} = - \bm \beta(\bm w)$, where the beta functions are listed in the Methods.
In the large-$N_f$ limit, the couplings admit an IR stable fixed point, $\bm w \rightarrow (\frac{6\pi^2 \epsilon}{N_f},0, \frac{4\pi^2 \epsilon}{N_f}, 0)$, establishing the existence of a deconfined fixed point for sufficiently large $N_f$.
To assess the fate of this deconfined fixed point at smaller $N_f$, we next track its evolution as $N_f$ is decreased and show that it is annihilated in a fixed-point collision, giving rise to pseudocritical scaling.

\begin{figure}
    \centering
    \includegraphics[width=1.0\linewidth]{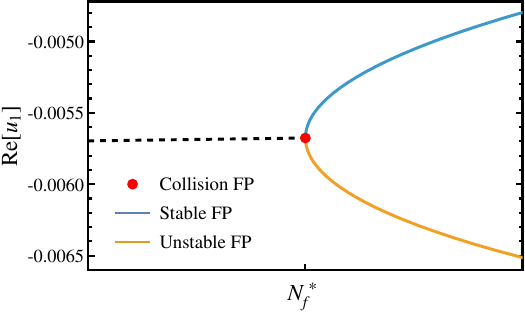}
    \caption{\textbf{Fixed-point collision near $N_f^\ast$ in the one-loop Sp(4) gauge--Higgs theory.}
    The real part of the fixed point value of $u_1$
    is plotted at $\epsilon=1$.
    After the fixed-point collision for $N_f<N_f^\ast$,
    the fixed points move to the complex plane.
    The red point denotes the collision fixed point.
    }
    \label{fig:FP}
\end{figure}

\vspace{10pt}
\noindent{\bf Fixed-point Collision and Pseudocritical Scaling}\\
As shown in Fig.~\ref{fig:FP}, the IR-stable fixed point collides with an unstable one at $N_f^\ast$~\footnote{Our one-loop calculation gives $N_f^\ast = 620.1 + \mathcal O (\epsilon)$. However, it is well-known that the one-loop results significantly overestimate the critical flavor at $\epsilon =1$~\cite{bonati2024charged,ihrig2019abelian}. 
Our analysis of the collision fixed point offers a possible interpretation of the DQMC scaling if $N_f^\ast$ lies sufficiently close to the physical value $N_f=4$ at $\epsilon=1$.
}; below this value the two fixed points annihilate and continue into the complex plane as a conjugate pair. 
At the collision, the fixed-point couplings are $\bm w^\ast  \approx \epsilon (0.107 ,  -0.006, 0.063, -0.006)$. 
This collision fixed point plays a central role in the ensuing pseudocritical behavior~\cite{nauenberg1980singularities,cardy1980scaling,gorbenko2018walking}. 
At $N_f = N_f^\ast$, one eigendirection of the linearized RG flow becomes marginal as an eigenvalue of the stability matrix vanishes. 
Denote the left eigenvector along the marginal direction as $\bm e_l$.
By projecting the couplings onto this direction, $s = \bm e_l \cdot (\bm w - \bm w^\ast)$, we find that the RG flow in the vicinity of the collision fixed point reduces to an effective one-dimensional equation of the form $\frac{{\rm d}s}{{\rm d}l} =  c_2 s^2 - c_0 \delta$, where $\delta = N_f - N_f^\ast$ and constants $c_{0,2} >0$ are given in Methods.   
For $\delta <0$, the fixed-point solutions move off the real axis, with $s \propto \pm i \sqrt{|\delta|}$, signaling that the real fixed points have annihilated into a complex-conjugate pair. 
Nevertheless, when 
$|\delta|$ is small the RG flow can remain close to the collision point for a parametrically long ``walking'' time:  
$l_{\rm walking} \propto \frac1{\sqrt{|\delta|}}$. 
As a result, the system exhibits pseudocritical scaling governed by the collision fixed point, with corrections controlled by the distance to the complex fixed points, $\mathcal O(\sqrt {|\delta|})$.

Our DQMC data in Fig.~\ref{fig:numeric2}(b,c) exhibit a slow, size-dependent drift of the effective scaling exponents. 
If, in three dimensions, $d=3$ ($\epsilon=1$), the critical value $N_f^\ast$ is close to the physical $N_f = 4$, then $\delta = N_f - N_f^\ast$ can be small, and the observed scaling may arise from a walking RG trajectory governed by the nearby collision fixed point. 
Motivated by this possibility, we evaluate scaling dimensions of gauge-invariant operators at the collision fixed point, 
$\eta^\ast \approx 0.91 $ and $ \nu^\ast \approx 0.67$,  with details provided in Methods. 
Hence, these collision-point estimates, shown by the dashed blue lines in Fig.~\ref{fig:numeric2}(b,c), motivate Sp(4) gauge-Higgs pseudocriticality as a possible interpretation of the observed scaling.

A microscopic route from the electronic Hubbard model to the Sp(4) gauge-Higgs description is suggested by a parton construction in which the electron fractionalizes into a bosonic matrix field and a charge-neutral fermion, generating an emergent $\mathrm{U}(4)\simeq[\mathrm{SU}(4)\times \mathrm{U}(1)]/\mathbb{Z}_4$ gauge redundancy~\cite{you2018symmetric}. 
In the charge-$2e$ phase, condensation of the matrix field locks flavor and gauge indices, fully Higgsing the gauge structure and recombining the fermionic parton into the physical electron. 
Away from the Higgs phase, a natural intermediate description is a parton Fermi surface coupled to both SU(4) gauge bosons and a U(1) gauge field. 
Crucially, the non-Abelian gauge fluctuations favor parton pairing~\cite{son1999superconductivity,zou2020deconfined}, and when this attraction dominates it destabilizes the parton Fermi surface. 
The resulting parton pairing remains finite through the transition, both maintaining a gap for single-electron excitations and reducing the emergent gauge structure to the subgroup preserving $\Omega$, namely Sp(4), leading to an Sp(4) gauge-Higgs critical theory as the long-wavelength description (see SI for a detailed discussion).

\section{Discussion}
Our numerically exact, large-scale DQMC calculations establish charge-$4e$ SC in the attractive SU(4) Hubbard model as an intrinsic zero-temperature phase. The results uncover a direct transition to a conventional charge-$2e$ superconductor in which single electrons remain gapped. Pairing correlations exhibit an anomalously large effective scaling dimension and systematic finite-size drift. These numerical observations motivate an Sp(4) gauge-Higgs theory whose generalized one-loop RG flow exhibits a fixed-point collision and pseudocritical behavior.
By linking the charge-$4e$--charge-$2e$ transition to non-Abelian gauge pseudocriticality, this result offers a possible interpretation of the numerical observations and a framework for investigating superconducting criticality beyond the conventional Landau description.
In the SI, we provide additional numerical results for the quarter filling, which also exhibit a large effective pairing anomalous dimension, indicating that this finite-size feature is not restricted to the $1/8$ filling.
Moreover, enhanced density correlations at quarter filling also suggest a competing CDW order; its precise phase boundary and possible coexistence with quartet coherence are left for future study.

More broadly, our work provides a theoretical framework for interpreting and guiding future searches for higher-charge SC in platforms where attractive interactions and SU(4) symmetries can be engineered. 
Looking ahead, extending our framework across fillings, lattice geometries, and SU(N) generalizations, and connecting to experimentally relevant multicomponent platforms will reveal a wider landscape of higher-charge superconductors and unconventional superconducting criticality beyond the Landau paradigm.
For instance, shielded ultracold molecules in optical lattices offer a plausible route to realizing an attractive SU(4) Hubbard model, which remains difficult to access in existing atomic platforms~\cite{bohn2017cold}. 
These systems provide a manifold of nearly degenerate internal states with interactions that are approximately spin independent, while also allowing the scattering length to be tuned through zero and into the attractive regime~\cite{mukherjee2025sun}. 
This would provide a natural cold-molecule platform for the attractive SU(4) Hubbard model and, consequently, a promising setting in which to search for quartet superfluidity, the neutral-matter analogue of charge-$4e$ order. 

In summary, our results connect zero-temperature charge-$4e$ superconductivity with an unconventional transition to the charge-$2e$ state, for which emergent non-Abelian gauge criticality offers a possible interpretation.

{\it Note added}: During the finalization of this manuscript, we became aware of a related work by S.-H. Shi, Z.-Z. Wu, J. Hu and Z.-X. Li, which appeared on arXiv at the same time as our work and reported primary charge-4e superconductivity in a doped SU(4) Su--Schrieffer--Heeger model~\cite{shi2026high}. 
Their study focuses on the finite-temperature phase diagram, including a charge-$4e$ Berezinskii--Kosterlitz--Thouless transition and pseudogap behavior. 
By contrast, our work focuses on the zero-temperature charge-$2e$--charge-$4e$ transition in the attractive SU(4) Hubbard model and its possible interpretation in terms of deconfined pseudocriticality.

\vspace{10pt}
\noindent \textbf{Acknowledgments} We would like to thank Zhi-Qiang Gao, Congjun Wu, Hui Yang, Yi-Zhuang You, Yan-Qi Wang, Liujun Zou and Hong Yao for helpful discussions.

\noindent \textbf{Author contributions} 
Z.-Q.W. performed the numerical simulations and data analysis. H.J. carried out the theoretical calculations. X.Z. contributed to the strong-coupling analysis. Z.-Q.W. and S.Z. developed the exact bridge link approach. S.-K.J. proposed the theoretical framework. S.Z. and S.-K.J. guided the research. All authors discussed the results and contributed to the writing of the manuscript.

\noindent \textbf{Funding Statement}
The Flatiron Institute is a division of the Simons Foundation. 

\noindent\textbf{Competing interests}
The authors declare no competing interests.

\noindent\textbf{Data availability}
The data supporting the findings of this study, including the data underlying all figures in the main text and Supplementary Information, are publicly available at
\url{https://github.com/Zhouquan-Wan/su4-charge4e-data}.

\noindent\textbf{Code availability}
The code used for data analysis is available in the GitHub repository linked above. The simulation code is available from the corresponding author upon reasonable request.

\bibliography{references}

\newpage
\clearpage

\section*{Methods}

\noindent{\bf DQMC Algorithm.}
The SU(4) Hubbard model defined in Eq.~\eqref{eq:hamiltonian} reads
\eq{
    \hat H= -t \sum_{a,\left<ij\right>}({c_i^a}^\dagger c_j^a+\text{h.c.}) -U \sum_{i}(\sum_{a}\hat n_i^a - 2)^2 \equiv \hat K + \hat V\,,
}{}
where $\hat K$ and $\hat V$ denote the kinetic and interaction terms, respectively.
To access ground-state properties, we employ the projector formalism, in which the ground state is obtained via imaginary-time evolution,
\begin{equation}
    \ket{\psi_G}\propto\lim_{\Theta\rightarrow \infty} e^{-\Theta \hat H}\ket{\psi_T}\,,
\end{equation}
starting from a trial wavefunction $\ket{\psi_T}$.

Within the projector DQMC framework, we discretize imaginary time via a Suzuki-Trotter decomposition,
\begin{equation}
e^{-\Theta \hat H}=\prod_{l=1}^{\Theta/\Delta\tau} e^{-\Delta\tau\hat K/2}  e^{-\Delta\tau\hat V}  e^{-\Delta\tau\hat K/2}+O(\Delta\tau^2)\,.
\end{equation}
The interaction term is further decoupled using a discrete Hubbard--Stratonovich (HS) transformation:
\eq{
e^{{\Delta \tau U} \left(\sum_a\hat n^a_i-2\right)^2} = \frac{1}{4}\sum_{s=\pm 1,\pm 2} \gamma(s) e^{\eta(s) \left(\sum_a\hat n^a_i-2\right)}\,,
}{}
with 
\eq{
\begin{split}
&\eta(\pm 1) = \pm \text{arccosh}\left((x+2x^3+x^5-(x^2-1)y)/4\right)\,,\\
&\eta(\pm 2)= \pm \text{arccosh}\left((x+2x^3+x^5+(x^2-1)y)/4\right)\,,\\
&\gamma(\pm 1) = 1 + x  (3+ x^2) / y\,,\\
&\gamma(\pm 2) = 1 - x  (3+ x^2) / y\,,\\
&x = e^{U\Delta\tau} \,, \quad y = \sqrt{8+x^2(3+x^2)^2}\,. \nonumber
\end{split} 
}{}

After the HS transformation, the imaginary-time propagator becomes fermionic Gaussian operators for any configuration $\{\mathbf{s}_l\}$.
As a result, if the trial wavefunction $\ket{\psi_T}$ is chosen as a Slater determinant, the propagated state remains a Slater determinant. The projected ground state can therefore be written as a superposition of Slater determinants, $\ket{\psi_G} \simeq \sum_{\{\mathbf{s}_l\}}\ket{\psi(\{\mathbf{s}_l\})}$, where $\ket{\psi(\{\mathbf{s}_l\})}$ denotes the state obtained by propagating $\ket{\psi_T}$ under the auxiliary-field configuration $\{\mathbf{s}_l\}$.

Because the propagator factorizes into four identical flavor sectors, we can also choose the trial wavefunction in the form $\ket{\psi_T} = \bigotimes_{a=1}^4 \ket{\psi_T^{(a)}}$, with identical Slater determinants $\ket{\psi_T^{(a)}}$. 
The propagated state then preserves the same factorized structure, $\ket{\psi(\{\mathbf{s}_l\})}=\bigotimes_{a=1}^4 \ket{\phi^{(a)}(\{\mathbf{s}_l\})}$, where the four flavor components are identical.
In the following, we will omit the flavor index $a$.

The Monte Carlo weight for a given auxiliary-field configuration is then given by the overlap between the left- and right-propagated states,
\begin{equation}
    W(\mathbf{s}) = \braket{\phi(\{\mathbf{s}^L_l\}) \mid \phi(\{\mathbf{s}^R_l\})}^4\,.
\end{equation}
Namely, the total weight is the fourth power of the single-flavor determinant (real) and is therefore non-negative. The simulation is thus free of the sign problem.

Ground-state expectation values are evaluated as Monte Carlo averages over auxiliary-field configurations,
\begin{equation}
    \langle \hat O \rangle
    =
    \left< \frac{
    \braket{\psi(\{\mathbf{s}^L_l\})|\hat O|\psi(\{\mathbf{s}^R_l\})}}{\braket{\psi(\{\mathbf{s}^L_l\})|\psi(\{\mathbf{s}^R_l\})}}\right>_{\mathbf{s}}\equiv
    \left< O_\text{loc}\right>_{\mathbf{s}}\,,
\end{equation}
where $\braket{\cdot}_{\mathbf{s}}$ denotes the Monte Carlo average over auxiliary-field configurations sampled with weight $W(\mathbf{s})$.
Since the propagated states are Slater determinants and remain factorized into four identical flavor sectors, local observables can be evaluated from the single-flavor Green's function $G_{ij}(\mathbf{s})$, defined as the local expectation value of $c_i^a {c_j^a}^\dagger$, using Wick's theorem.

All simulations in the main text are performed with $2\Theta = 50 + L$ and $\Delta \tau = 0.1$ unless otherwise specified. We have checked that increasing $\Theta $ does not change the results within statistical errors. The choice of $\Delta\tau$ is not expected to affect the critical behavior at the transition.
The leading computational cost per Monte Carlo sweep scales as $\mathcal{O}(M N_s^3)$, where $N_s=L^2$ and $M=2\Theta/\Delta\tau$. 
Each Monte Carlo sweep consists of one bidirectional traversal of the full imaginary-time projection interval, supplemented by ten additional bidirectional traversals restricted to the central measurement window to improve the statistics of the bridge measurements.
The simulations underlying the main-text results required approximately $7.4\times10^5$ CPU core-hours, or $1.25\times10^6$, including additional runs and convergence tests.
For example, at $L=52$, a 2,500-sweep Markov chain took approximately 152 hours on eight CPU cores.
We ran 40 independent Markov chains for each interaction strength.

\vspace{10pt}
\noindent{\bf Infinite variance and exact bridge link method.}
Although the DQMC simulation is free from the sign problem, the factorized determinant structure implies that the Monte Carlo estimator of the charge-\(4e\) correlation function suffers from an infinite-variance problem in the charge-\(4e\) phase.
Specifically, correlations in DQMC are computed from the single-flavor Green's function as
\eq{
\begin{split}
&\braket{\Delta_{2e}(i)\Delta_{2e}^\dagger(j)} = \braket{G_{ij}^2(\mathbf{s})}_{\mathbf{s}},\\
&\braket{\Delta_{4e}(i)\Delta_{4e}^\dagger(j)} = \braket{G_{ij}^4(\mathbf{s})}_{\mathbf{s}}.
\end{split}
}{}
The variance of the charge-$4e$ correlation is then given by $\braket{G_{ij}({\mathbf{s}})^8}_{\mathbf{s}}-\braket{G_{ij}({\mathbf{s}})^4}_{\mathbf{s}}^2$ which satisfies the inequality
\eq{
\frac{\braket{G_{ij}({\mathbf{s}})^8}_{\mathbf{s}}}
{\braket{G_{ij}({\mathbf{s}})^4}_{\mathbf{s}}^2}\geq
\frac{\braket{G_{ij}({\mathbf{s}})^4}_{\mathbf{s}}}{\braket{G_{ij}({\mathbf{s}})^2}_{\mathbf{s}}^2}.
}{}
In the charge-$4e$ phase, when $i$ and $j$ are separated by the maximum distance on the lattice, the expectation values behave as $\braket{G_{ij}({\mathbf{s}})^2}_{\mathbf{s}}\rightarrow 0$, $\braket{G_{ij}({\mathbf{s}})^4}_{\mathbf{s}}\rightarrow \text{const.}$ in the thermodynamic limit.
As a result, the right-hand side of the above inequality diverges, which implies that the variance of the charge-$4e$ estimator also diverges. Direct measurements of charge-$4e$ correlations therefore suffer from an infinite-variance problem.

\begin{figure}
    \centering
    \includegraphics[width=1.0\linewidth]{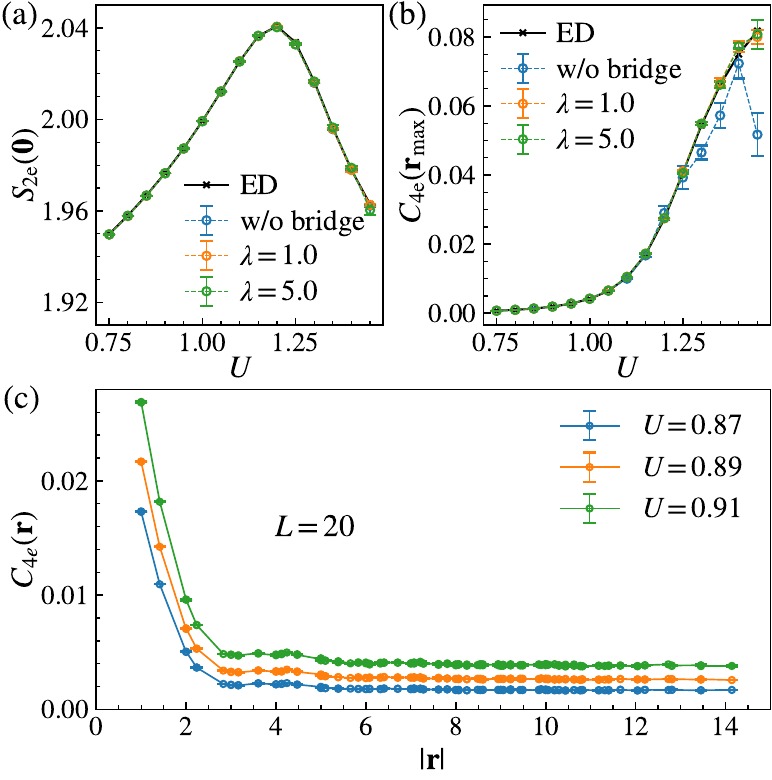}
    \caption{\textbf{Bridge-link benchmark and real-space charge-$4e$ correlations.} (a,b) Benchmark of the bridge link method for a $4\times4$ system with one electron per flavor. 
    (a) Charge-$2e$ structure factor computed using different methods. 
    (b) Charge-$4e$ correlation function at the largest separation computed using different methods.  
    The bridge link method removes the bias in the charge-$4e$ correlation, while all methods give consistent results for the charge-$2e$ structure factor. 
    For each point, $8\times 10^5$ samples are collected for all methods.
    (c) The charge-$4e$ correlation $C_{4e}(\mathbf{r})$ for a system with $L=20$, obtained using the exact bridge link method. 
    As $U$ is tuned across the phase transition, the correlations all exhibit a flat plateau at large distances, consistent with long-range charge-$4e$ order.
    The bridge link operator $\hat \Lambda(\lambda,\mathbf{r})$ with $\lambda=5$ is applied at separations corresponding to $\mathbf{r}$.
    }
    \label{fig:benchmark}
\end{figure}

To address the infinite-variance problem, we apply the exact bridge link method. 
The central idea is to eliminate the zeros of the sampling weight by inserting a bridge operator at the midpoint of the propagators.
Explicitly, we use the following bridge operator
\eq{
\hat \Lambda(\lambda,\mathbf{r}) = \frac 1 {L^2} \sum_{i}
\left( e^{\lambda \sum_a c^{a\dagger}_i c^a_{i+\mathbf{r}}}
+e^{\lambda \sum_a c^{a\dagger}_{i+\mathbf{r}} c^a_i}\right)\,,
}{SUNbridge}
where $\lambda \in \mathbb{R}$, and $\mathbf{r}$ denotes the separation.
As shown in Ref.~\cite{wan2025bridge}, it can eliminate the divergence of the Monte Carlo estimator for the charge-$4e$ correlation $C_{4e}(\mathbf{r})$ at separation $\mathbf{r}$, thereby solving the infinite-variance problem. 

Next, we benchmark the method against exact diagonalization (ED) results in a small system. 
As shown in Fig.~\ref{fig:benchmark}(a,b), standard DQMC without the bridge link yields charge-$4e$ correlations that deviate significantly from the exact results at large interaction strength. 
The bridge link method resolves this problem, while preserving the correctness of the charge-$2e$ structure factor.
In Fig.~\ref{fig:benchmark}(c), we compute the charge-$4e$ correlation function for a larger system with $L=20$. 
To evaluate correlations at each distance, the bridge operator is applied at the corresponding separation. 
The resulting correlations exhibit a clear plateau at large distances, consistent with long-range charge-$4e$ order.

To reduce computational cost, in Fig.~\ref{fig:numeric1} we evaluate only $C_{4e}(\mathbf{r}_{\rm max})$ using $\hat{\Lambda}(\lambda,\mathbf{r}_{\rm max})$, which is sufficient to demonstrate long-range charge-$4e$ order. 
Throughout this work, we set $\lambda=5$ unless otherwise specified.

\vspace{10pt}
\noindent
{\bf Renormalization Group Analysis for the Sp(4) Gauge-Higgs Model.}
The critical theory can be generalized to Sp(4)-Higgs theory with SU($N_f$) global symmetry. 
The gauge redundancy can be fixed by introducing a ghost fermion field $c$ in the adjoint representation. 
Explicitly, we have 
\begin{eqnarray}
    \mathcal L &=& \Tr[(\partial_\mu Z - i g Z A_\mu)(\partial_\mu Z^\dag + i g A_\mu Z^\dag)] + \frac14 F_{\mu\nu}^i F_{\mu\nu}^i \nn \\
    && + \partial_\mu \bar c^{i}  \partial_\mu c^i - g f^{ijk} (\partial_\mu \bar c^{i}) A_\mu^j c^k +  u_1 [\Tr (Z Z^\dag)]^2 \nn \\
    &&  + u_2 \Tr[(Z Z^\dag)^2]  + v \Tr[ |Z \Omega Z^T |^2 ] \,.
\end{eqnarray}
Here, $Z_{a\alpha}$ is the Higgs boson carrying a global SU($N_f$) fundamental representation ($a$) and a gauge Sp(4) fundamental representation ($\alpha$).
The trace is over the SU($N_f$) fundamental. 
$c^i$ is the ghost fermion field, with the gauge Sp(4) adjoint $i$.
Note that it is not the electron operator. 
$A_\mu = A_\mu^i T^i$ is the gauge field, where $T^i$ is the Sp(4) generator, satisfying 
\begin{eqnarray}
    [T^i,T^j] = i f^{ijk} T^k\,, \quad T^j \Omega + \Omega (T^j)^T = 0 \,.
\end{eqnarray}
Our convention for the Sp(4) group is $\Tr[T^i T^j] = 4 \delta^{ij} $. 
Finally, $F_{\mu \nu}^i = \partial_\mu A^i_{\nu} - \partial_\nu A_\mu^i - gf^{ijk} A^j_\mu A^k_\nu$ is the field strength tensor.

The one-loop RG calculation for the non-Abelian gauge-Higgs model is standard.
The propagators, which are the building blocks for the Feynman diagrams, are
\begin{eqnarray}
    \left< Z_{a,\alpha}(p) Z_{b,\beta}^\ast(p') \right> &=& \delta_{\alpha\beta} \delta_{ab}  \delta(p-p') \frac1{p^2} \,,  \\
    \left< c^i(p) \bar c^{j}(p') \right> &=& \delta^{ij}  \delta(p-p') \frac1{p^2} \,, \\
    \left< A_\mu^i(p) A_\nu^{j}(p') \right> &=& \delta_{\mu\nu} \delta^{ij}  \delta(p+p') \frac1{p^2} \,.
\end{eqnarray}
The one-loop Feynman diagrams and corresponding renormalization factors are given in the ``One-loop renormalization group'' section of the Supplementary Information.
We define the field anomalous dimensions by $\eta_X=\mu\,{\rm d}\ln Z_X^{1/2}/{\rm d}\mu$, with $X=Z,A,c$, where $Z_X$ denotes the renormalization factor, and the derivative is taken at fixed bare parameters.
The resulting anomalous dimensions are:
\begin{equation}
\begin{split}
    & \eta_Z = - \frac{5g^2}{4\pi^2}\,, \quad \eta_c = - \frac{3g^2}{4\pi^2} \,, \\
    & \eta_A = - \frac{5g^2}{2\pi^2} + N_f \frac{g^2}{12\pi^2} \,.
\end{split}
\end{equation}

The one-loop RG equation for the gauge coupling strength is
\begin{eqnarray} \label{eq:RG_g}
    \frac{{\rm d}g}{{\rm d}l} &=& \frac{\epsilon}2 g - \frac{1}{12\pi^2}(N_f-66) g^3 \,.
\end{eqnarray}

Finally, the one-loop corrections to quartic Higgs boson interactions result from the boson self-interaction and the boson-gauge couplings.
The resulting RG equations are
\begin{eqnarray} \label{eq:RG_u}
    \frac{{\rm d}u_1}{{\rm d}l} &=& \epsilon u_1
    +\left(\frac{15g^2}{2\pi^2}
    -\frac{(N_f+4)u_2}{2\pi^2}
    -\frac{(N_f-1)v}{2\pi^2}\right)u_1 \nn\\
    &&-\frac{N_f+1}{\pi^2}u_1^2
    -\frac{3g^4}{2\pi^2}
    -\frac{3u_2^2}{4\pi^2}
    -\frac{u_2v}{2\pi^2}
    -\frac{v^2}{4\pi^2}\,,
     \\
    \frac{{\rm d}u_2}{{\rm d}l} &=& \epsilon u_2
    +\left(\frac{15g^2}{2\pi^2}
    -\frac{3u_1}{2\pi^2}
    +\frac{v}{2\pi^2}\right)u_2 \nn\\
    &&-\frac{9g^4}{2\pi^2}
    -\frac{(N_f+4)u_2^2}{4\pi^2}
    -\frac{(N_f-2)v^2}{4\pi^2}\,, \\
    \frac{{\rm d}v}{{\rm d}l} &=& \epsilon v-\frac{v^2}{\pi^2}+\left(\frac{15g^2}{2\pi^2}-\frac{3u_1}{2\pi^2}-\frac{(N_f-2)u_2}{2\pi^2}\right)v\nn\\&&-\frac{9g^4}{2\pi^2}\,.
\end{eqnarray}

We also calculate the scaling dimension of the gauge-invariant quantities at the critical point. 
The scaling dimension of the charge 2e order, $ Z \Omega Z^T$, is
\begin{eqnarray}
    d_{2e} = d-2 - \frac{15}{4\pi^2}g^2 - \frac{1}{4\pi^2}(-u_1 + u_2 - 4 v)\,.
\end{eqnarray}
And the scaling dimension for the mass singlet, $\Tr[Z Z^\dag]$, is 
\begin{equation}
\begin{split}
    d_m =& \,\,d-2-\frac{15g^2}{4\pi^2}+\frac{(1+4N_f) u_1}{4\pi^2} \\
    &+\frac{(4+N_f)u_2}{4\pi^2}+\frac{(N_f-1)v}{4\pi^2} \,. 
\end{split}
\end{equation}

For convenience, we introduce the convention $\frac{{\rm d} \bm w}{{\rm d}l} = - \bm \beta(\bm w)$, with $\bm w = (g^2, u_1, u_2, v)$, and define the Jacobian $J_{ij} = - \frac{\partial \beta_i}{\partial w_j}$. 
It is straightforward to show that in the leading order of $1/N_f$, there exists an IR stable fixed point given by $\bm w \rightarrow (\frac{6\pi^2 \epsilon}{N_f},0, \frac{4\pi^2 \epsilon}{N_f}, 0)$.   

We track the IR stable fixed point from the large-$N_f$ limit when $N_f$ decreases. 
Eventually, the IR stable fixed point collides with an unstable fixed point at a critical flavor $N_f^\ast$, which can be determined by $\bm \beta = 0$ and $\det J =0$, 
$N_f^*=620.10$ and the collision fixed point 
	\begin{align}
		{\bm w}^*\approx (0.107\epsilon ,  -0.006\epsilon, 0.063\epsilon, -0.006\epsilon)\, .
	\end{align}
At $N_f = N_f^\ast$, the following eigendirection of the linearized RG flow becomes marginal as an eigenvalue of the stability matrix vanishes:
\begin{eqnarray}
	\bm e = (0, -0.999, 0.011, 0.001)\,.
\end{eqnarray}
Let $\bm e_l$ denote the corresponding left zero eigenvector, satisfying $\bm e_l\cdot\bm J=0$.
Defining $s=\bm e_l\cdot(\bm w-\bm w^\ast)$, the RG flow in the vicinity of the collision fixed point reduces to an effective one-dimensional equation of the form
\begin{eqnarray}
    \frac{{\rm d}s}{{\rm d}l} =  c_2 s^2 - c_0  \delta + O(s^3,s\delta,\delta^2) \,,
\end{eqnarray}
where $\delta = N_f - N_f^\ast$ and $c_2 =62.7$, $c_0 = 8.0 \times 10^{-6}$.
For $\delta<0$, i.e., $N_f$ less than $N_f^\ast$, the two fixed points become a conjugate pair in the complex plane, $\bm w = {\bm w}^\ast \pm i \sqrt{\frac{c_0}{c_2} |\delta|} {\bm e} +\mathcal{O}(\delta)$. 
Nevertheless, when $|\delta|$ is small the RG flow can remain close to the collision point for a parametrically long RG time.
In particular, the slow flow along the nearly marginal direction produces 
\begin{equation}
    l_{\rm walking} = \int \frac{{\rm d}s}{c_2 s^2 - c_0 \delta} \propto \frac1{\sqrt{|\delta|}} \,.
\end{equation} 
As a result, the generalized theory exhibits pseudocritical scaling governed by the collision fixed point, with corrections controlled by the distance to the complex fixed points, $\mathcal O(\sqrt {|\delta|})$.
In particular, the drifting critical exponent is approximately 
\begin{eqnarray}
    \eta = \eta^\ast + \mathcal{O}(\sqrt {|\delta|}) \,, \quad \nu = \nu^\ast + \mathcal{O}(\sqrt {|\delta|}) \,,
\end{eqnarray}
where $\eta^\ast$ and $\nu^\ast$ are critical exponents at the collision fixed point.
To one-loop order, we obtain
\begin{align}
    \eta^\ast
    &=2-1.0859\epsilon+O(\epsilon^2),\\
    (\nu^\ast)^{-1}
    &=2-0.4994\epsilon+O(\epsilon^2).
\end{align}
Setting $\epsilon=1$, we have $\eta^\ast\approx0.91$ and $ \nu^\ast\approx0.67$.

\setcounter{secnumdepth}{3}
\setcounter{equation}{0}
\setcounter{figure}{0}
\renewcommand{\theequation}{S\arabic{equation}}
\renewcommand{\thefigure}{S\arabic{figure}}
\renewcommand\figurename{Supplementary Figure}
\renewcommand\tablename{Supplementary Table}

\clearpage
\section*{Supplementary Information}
\section*{Strong Coupling Analysis for the SU(4) Hubbard Model}

We perform a degenerate perturbation analysis for the SU(4) Hubbard model in the strong coupling limit. 
The Hamiltonian can be decomposed as $H = H_0 + V$, where the unperturbed term corresponds to the on-site Hubbard interaction, $H_0=-U\sum_{i}\left(\sum_{a}\hat n_{i}^a-2\right)^2$, and the perturbation describes the nearest-neighbor (NN) hopping term, $V=-t\sum_{a,\langle ij\rangle}\left({c_{i}^a}^{\dagger} c_{j}^a+\text{h.c.}\right)$. 
We introduce the projection operator $P$ to the low-energy subspace of $H_0$, and its complement, $Q = 1 - P$:
\begin{eqnarray}
    P H_0 P = E_P P\,, \quad  PH_0 Q  = QH_0 P =0 \,,
\end{eqnarray}
where $E_P$ denotes the ground state energy of $H_0$. 
Applying degenerate perturbation theory, we can obtain the Brillouin-Wigner (BW) effective Hamiltonian acting on the degenerate low-energy subspace. Omitting the common unperturbed energy, the effective Hamiltonian up to fourth order reads:
\begin{eqnarray}
    H_{\rm eff} &=& P (V R) V P + P (VR)^3 V P \nonumber \\
    &&  - P V R^2 V P V RV P \,,
    \label{eq:hameff}
\end{eqnarray}
where $R = Q(E_P - Q H_0 Q)^{-1}Q$ is the propagator.

The unperturbed Hamiltonian $H_0$ has two locally degenerate ground states, denoted by $\ket{0}$ and $\ket{4}$. 
These correspond to the empty and fully occupied states at each site, respectively.
Therefore, within the low-energy space, the effective Hamiltonian can be expressed in terms of hard-core boson operators $b_i$, $b_i^\dag$ and the number operator $n_i = b_i^\dag b_i$. 
The NN hopping term $V$ serves as a perturbation that transfers a single electron between adjacent sites. 
Up to fourth order, the following processes contribute to the effective Hamiltonian:
\begin{align}
    &\ket{4,0}\to\ket{3,1}\to\ket{4,0},\nonumber\\
    &\ket{4,0}\to\ket{3,1}\to\ket{2,2}\to\ket{1,3}\to\ket{0,4},\nonumber\\
    &\ket{4,0}\to\ket{3,1}\to\ket{4,0}\to\ket{3,1}\to\ket{4,0},\nonumber\\
    &\ket{4,0}\to\ket{3,1}\to\ket{2,2}\to\ket{3,1}\to\ket{4,0},\nonumber\\
    &\ket{4,0,0}\to\ket{3,1,0}\to\ket{3,0,1}\to\ket{3,1,0}\to\ket{4,0,0},\nonumber\\
    &\ket{0,4,0}\to\ket{1,3,0}\to\ket{0,4,0}\to\ket{0,3,1}\to\ket{0,4,0},\nonumber\\
    &\ket{0,4,0}\to\ket{1,3,0}\to\ket{1,2,1}\to\ket{0,3,1}\to\ket{0,4,0},\nonumber\\
    &\ket{\begin{smallmatrix}
        4 & 0\\
        0 & 0
    \end{smallmatrix}}\to
    \ket{\begin{smallmatrix}
        3 & 1\\
        0 & 0
    \end{smallmatrix}}\to
    \ket{\begin{smallmatrix}
        3 & 0\\
        0 & 1
    \end{smallmatrix}}\to
    \ket{\begin{smallmatrix}
        3 & 0\\
        1 & 0
    \end{smallmatrix}}\to
    \ket{\begin{smallmatrix}
        4 & 0\\
        0 & 0
    \end{smallmatrix}},\nonumber\\
    &\ket{\begin{smallmatrix}
        4 & 4\\
        0 & 0
    \end{smallmatrix}}\to
    \ket{\begin{smallmatrix}
        4 & 3\\
        0 & 1
    \end{smallmatrix}}\to
    \ket{\begin{smallmatrix}
        4 & 3\\
        1 & 0
    \end{smallmatrix}}\to
    \ket{\begin{smallmatrix}
        3 & 4\\
        1 & 0
    \end{smallmatrix}}\to
    \ket{\begin{smallmatrix}
        4 & 4\\
        0 & 0
    \end{smallmatrix}},\nonumber\\
    &\ket{\begin{smallmatrix}
        4 & 4\\
        0 & 0
    \end{smallmatrix}}\to
    \ket{\begin{smallmatrix}
        4 & 3\\
        0 & 1
    \end{smallmatrix}}\to
    \ket{\begin{smallmatrix}
        3 & 4\\
        0 & 1
    \end{smallmatrix}}\to
    \ket{\begin{smallmatrix}
        3 & 4\\
        1 & 0
    \end{smallmatrix}}\to
    \ket{\begin{smallmatrix}
        4 & 4\\
        0 & 0
    \end{smallmatrix}},\nonumber\\
    &\ket{\begin{smallmatrix}
        4 & 0\\
        0 & 4
    \end{smallmatrix}}\to
    \ket{\begin{smallmatrix}
        3 & 1\\
        0 & 4
    \end{smallmatrix}}\to
    \ket{\begin{smallmatrix}
        3 & 1\\
        1 & 3
    \end{smallmatrix}}\to
    \ket{\begin{smallmatrix}
        4 & 1\\
        0 & 3
    \end{smallmatrix}}\to
    \ket{\begin{smallmatrix}
        4 & 0\\
        0 & 4
    \end{smallmatrix}},
\end{align}
where $\ket{i,j}$ ($\ket{i,j,k}$) denotes configurations with occupation on two (three) sites, respectively, while $\ket{\begin{smallmatrix}i & j\\k & l\end{smallmatrix}}$ represents the occupation pattern on the four sites of a plaquette.
The effective coupling between hard-core bosons can be extracted from the perturbative processes in Eq.~\eqref{eq:hameff}. For instance, the second-order process $\ket{4,0}\to\ket{3,1}\to\ket{4,0}$ leads to a density-density interaction of the form $t_1\sum_{\langle ij\rangle}\left[ n_{i}(1-n_{j})+(1-n_{i})n_{j}\right]$, with coupling strength $t_1=4\cdot(-t)\frac{1}{-6U}(-t)=-\frac{2t^2}{3U}$ from the contribution $PVRVP$.

By systematically accounting for all perturbative processes, we obtain the effective Hamiltonian,
\begin{align}
    H_{\rm {eff}} =&
    - t_b\sum_{\langle ij\rangle}\left( b_i^\dagger b_j + b_j^\dagger b_i \right)
    +V_1\sum_{\langle ij\rangle} n_i n_j
    +V_2\sum_{\langle ij\rangle_2} n_i n_j \nonumber\\
    &+V_3\sum_{\langle ij\rangle_3} n_i n_j
    +V_L\sum_{\langle ikj\rangle_L} n_i n_k n_j
    +V_p\sum_{\langle ijkl\rangle} n_i n_j n_k n_l \nonumber\\
    &+\mu\sum_i n_i \,.
\end{align}
Here, $V_2$ and $V_3$ denote the next-nearest and next-next-nearest neighbor density-density interactions, respectively. 
The term proportional to $V_L$ describes a three-site interaction along an L-shaped path $\langle ikj\rangle_L$, while $V_p$ denotes a four-site plaquette interaction. 
Each bond, L-shaped triplet, and elementary plaquette is counted once.
The coefficient $\mu$ is a perturbatively generated one-body energy shift. 
The effective coupling constants are summarized as:
\begin{align}
    t_b &= \frac{t^4}{12 U^3}\,, \quad V_1 = \frac{4t^2}{3 U}+ \frac{61t^4}{270 U^3}\,, \quad V_2 = \frac{34t^4}{135 U^3}\,, \nonumber \\
    V_3 &= \frac{7t^4}{135U^3}\,,\quad V_L = -\frac{10t^4}{27U^3}\,, \quad V_p = \frac{20t^4}{27U^3}\,, \nonumber\\
    \mu &= -\frac{8t^2}{3 U}-\frac{43t^4}{135U^3}\,.
\end{align}

\begin{figure}[t]
    \centering
    \includegraphics[width=1.0\linewidth]{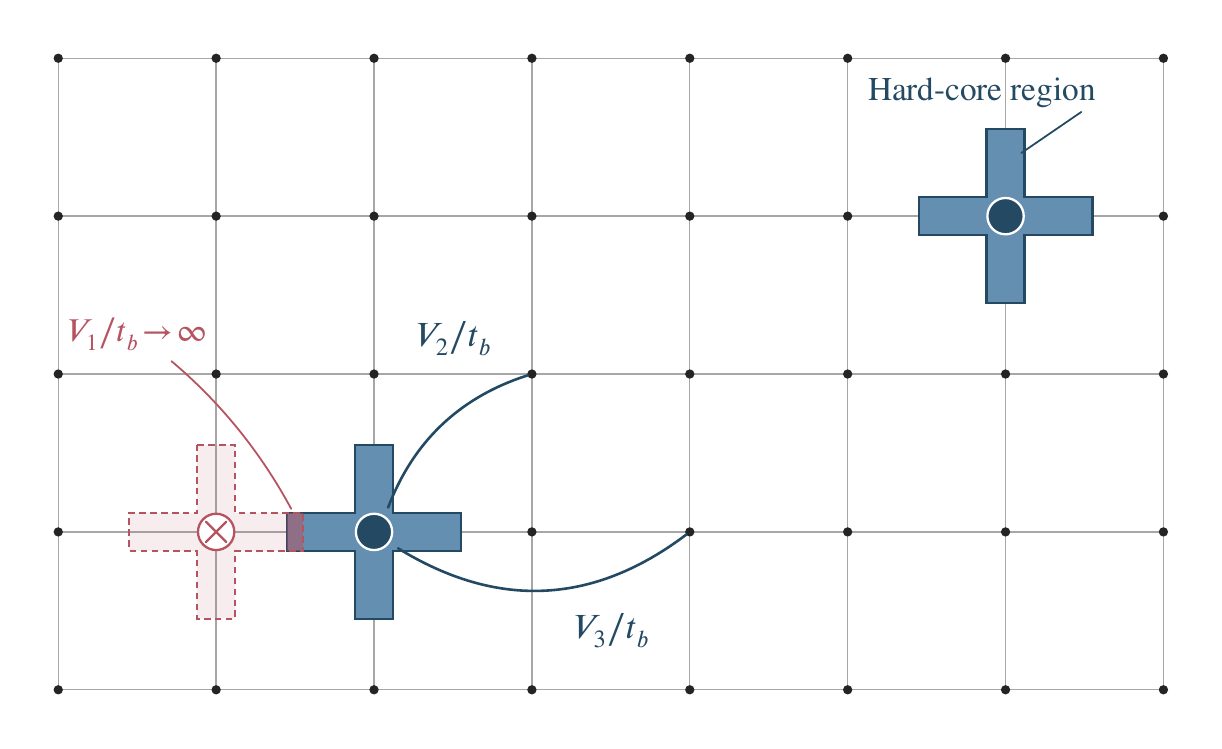}
    \caption{\textbf{Schematic of the constrained quartet-boson model in the strong-coupling limit.} Blue disks represent quartet bosons, with cross-shaped hard cores enforcing nearest-neighbor exclusion ($V_1/t_b\to \infty$). The red dashed cross illustrates a forbidden placement. The blue arcs labeled $V_2/t_b$ and $V_3/t_b$ indicate the diagonal and axial distance-two repulsions, respectively.}
    \label{fig:strong_coupling_limit}
\end{figure}

In the strong-coupling limit, $V_1/t_b=16(U/t)^2+ \mathcal O(1)$ diverges,
whereas the remaining fourth-order couplings remain finite in units of
$t_b$. For quartet densities $n_b<1/2$, the low-energy Hilbert space is
therefore restricted to configurations without occupied NN pairs.
For fillings above $1/2$, the same construction applies to holes after
a particle--hole transformation.
Defining $\mathcal P_0=\prod_{\langle ij\rangle}(1-n_i n_j)$,
the leading effective Hamiltonian at fixed boson number reduces, up to a
constant, to
\begin{align}
\frac{H_{\mathrm c}}{t_b}
=\mathcal P_0\bigg[&-\sum_{\langle ij\rangle}
(b_i^\dagger b_j+\mathrm{h.c.})\nonumber\\
&+\frac{136}{45}\sum_{\langle ij\rangle_2}n_i n_j
+\frac{28}{45}\sum_{\langle ij\rangle_3}n_i n_j
\bigg]\mathcal P_0.
\label{eq:strong_coupling_constrained}
\end{align}
The one-body energy shift is constant at fixed quartet number $N_b$, while the $V_L$ and $V_p$ terms vanish in this subspace because their configurations necessarily contain occupied NN pairs. The resulting constraint and the remaining diagonal and axial distance-two interactions are illustrated in Fig.~\ref{fig:strong_coupling_limit}.
The constrained model thus describes a lattice gas of quartet bosons with short-range repulsive interactions. 
In the dilute limit $n_b\to0$, we expect a zero-temperature superfluid phase of these bosons, corresponding to charge-$4e$ superconductivity.

At higher densities, these repulsions can also favor charge-density-wave (CDW) order. 
Classically, the maximum density at which occupied NN, diagonal, and axial distance-two pairs can all be avoided is $n_b=1/5$.
Our choice $n_b=1/8<1/5$ therefore allows such configurations and provides a cleaner setting for examining quartet coherence. 
Our additional calculations at $1/4$ filling show pronounced CDW signatures at larger interactions (Fig.~\ref{fig:quarter_density}), although residual sampling drift limits a controlled determination of the phase boundary. 
The strong-coupling expansion provides an asymptotic physical picture of the charge-$4e$ regime. 
However, we should note that its projected Hilbert space contains only empty sites and quartets, so this description does not fully capture the transition between charge-$2e$ and charge-$4e$ superconductivity.
Even if CDW order is favored in the asymptotic strong-coupling limit, charge-$4e$ superconductivity, or its coexistence with CDW order, may appear at intermediate coupling.
This competition and possible coexistence are interesting directions for further investigation.

\section*{Landau--Ginzburg--Wilson Theory}

In this section, we consider the potential Landau--Ginzburg--Wilson (LGW) theory for the transition from a charge-$4e$ superconducting state to a charge-$2e$ superconducting state. 
The charge-$4e$ state preserves the global SU(4) symmetry, while the charge-$2e$ state spontaneously breaks it down to Sp(4).  
More precisely, the charge-$2e$ superconducting order parameter is formed by a bilinear of SU(4) fundamental fermions.
The corresponding tensor product can be decomposed as
\begin{eqnarray}
    {\bm4} \otimes\bm 4 = \bm6_{\textbf{A}} \oplus \bm{10}_{\textbf{S}} \,.
\end{eqnarray}
where $\bm 6_{\textbf{A}}$ and ${\bm {10}}_{\textbf{S}}$ denote the antisymmetric and symmetric irreducible representations, respectively.
Hence, the order parameter transforms in the representation ${\bm6}_{\textbf{A}}$ as
\begin{eqnarray}
    \Delta \rightarrow U \Delta U^T \,, \quad U \in {\rm SU(4)}\,.
\end{eqnarray}
The residual symmetry is determined by the pattern of the order.  
Guided by the DQMC results, the order parameter $\Delta_{ab} = \left< c_a c_b \right>$ can be brought to an equal pairing between flavors one and three, as well as two and four.
Namely, the order parameter is proportional to the following matrix,
\begin{eqnarray} \label{eq:Omega}
    \Omega = \left(\begin{array}{cc}
        0 & I_{2} \\
        - I_2 & 0
    \end{array}\right)\,,
\end{eqnarray}
where $I_2$ is a $2\times 2$ identity matrix. 
The residual symmetry group follows the condition
\begin{eqnarray}
    U \Omega U^T = \Omega,
\end{eqnarray}
which is precisely the compact symplectic group Sp(4).

A direct transition from the charge-$4e$ phase to the charge-$2e$ phase is described by the coset theory SU(4)/Sp(4). 
To formulate the corresponding field theory, we first specify a representation of the SU(4) and Sp(4) generators.
The generators of Sp(4), denoted by $T^i$, satisfy the relation,
\begin{eqnarray}\label{eq:Sp4_algebra} 
    T^i \Omega + \Omega (T^i)^T = 0\,, \quad i=1,2,...,10\,. 
\end{eqnarray}
A convenient representation of the generators for Sp(4) can be constructed using a Clifford algebra. 
Consider the following representation for the Clifford algebra,
\begin{eqnarray}
    Y_a = \{\sigma^x, \tau^z \sigma^y, \tau^x \sigma^y, \tau^y \sigma^y, \sigma^z \},
\end{eqnarray}
which satisfy the anticommutation relations $\{Y_a, Y_b \} = 2 \delta_{ab}$. 
We can build ten Hermitian matrices from the commutators, $\frac1{2i} [Y_a, Y_b]$.
These matrices satisfy Eq.~\eqref{eq:Sp4_algebra}, and therefore furnish the generators of Sp(4).

Closely related, the generators of SU(4) are given by traceless $4 \times 4 $ Hermitian matrices.
A convenient basis is provided by the set $\{Y_a, T^i\}$, which together form the 15 generators of SU(4). 
This also suggests that $Y_a$ are the generators of the coset SU(4)/Sp(4). 

Now, we use the generators of SU(4)/Sp(4) to construct the coset field, 
\begin{eqnarray}
    M(x) = e^{i \pi_a(x) Y_a} \Omega e^{i \pi_b(x) Y_b^T}\,.
\end{eqnarray}
where $\pi_a(x)$ denote the real bosonic field describing long-wavelength fluctuations, and summation over $a$ and $b$ is implied.
The corresponding nonlinear sigma model (NLSM) is given by
\begin{eqnarray}
    \mathcal L = \frac1{8g} \Tr[\partial_\mu M^\dag(x) \partial^\mu M(x)] \,,
\end{eqnarray}
where $\partial_\mu$ is the spacetime derivative, and $M^\dag(x) M(x) = I$. 

One alternative parametrization of the order parameter can be obtained by expressing it in terms of six real fields $\phi_i$, giving the matrix
\begin{eqnarray}
    \left( \begin{array}{cccc} 0 & \phi_1 + i \phi_2 & \phi_3 + i \phi_4 & \phi_5 + i \phi_6 \\
    -\phi_1 - i \phi_2 & 0 &  -\phi_5 + i \phi_6 &\phi_3 - i \phi_4 \\
    -\phi_3 - i \phi_4 &  \phi_5 - i \phi_6 & 0 & -\phi_1 + i \phi_2 \\
    - \phi_5 - i \phi_6 & -\phi_3 + i \phi_4 & \phi_1 - i \phi_2  & 0 \end{array}\right) \,. \nn \\ 
\end{eqnarray}
With this parametrization, the theory can be simplified to an SO(6)/SO(5) NLSM,
\begin{eqnarray}
    \mathcal L = \frac1{2g }   \partial_\mu \phi_i \partial^\mu \phi_i\,, \qquad \sum_{i = 1}^6 \phi_i^2 = 1 \,,
\end{eqnarray}
reflecting the isomorphism SU(4)/Sp(4) $\cong$ SO(6)/SO(5). 

\section*{One-loop renormalization group}
\label{sec:one_loop_renormalization_constants}

\begin{figure}
    \centering
    \includegraphics[width=1\linewidth]{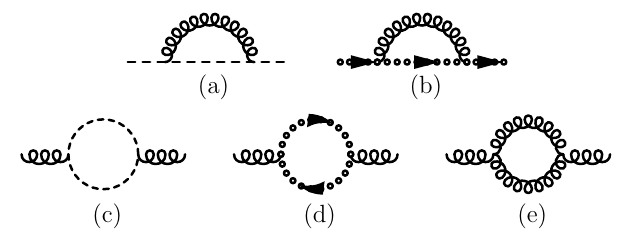}
    \caption{\textbf{One-loop self-energy diagrams.} Panels (a), (b), and (c--e) show the self-energy diagrams for the Higgs boson, ghost field, and gauge boson, respectively. 
    Dashed lines, dotted arrows, and curly lines denote the propagators for the Higgs boson, the ghost field, and the gauge boson, respectively.}
    \label{fig:feyn_2pt}
\end{figure}

We use dimensional regularization in $d=4-\epsilon$, minimal subtraction, and Feynman gauge, with the action and generator normalization specified in the Methods. 
The following results refer to the three-coupling quartic sector $(u_1,u_2,v)$, with the symplectic invariant defined as $\operatorname{Tr}(BB^\dagger)$, where $B=Z\Omega Z^T$.
Bare and renormalized fields and the gauge coupling are related by
\begin{equation}
 X_0=Z_X^{1/2}X_R,\qquad
 g_0=\mu^{\epsilon/2}Z_g g_R,
 \label{eq:rg_bare_fields}
\end{equation}
where $X=Z,A,c,\bar c$ and $Z_{\bar c}=Z_c$; internal indices are suppressed. 
All couplings below are renormalized, with the subscript $R$ omitted. 
The one-loop self-energy diagrams are shown in Fig.~\ref{fig:feyn_2pt}. The corresponding field-renormalization factors are
\begin{align}
 Z_Z&=1+\frac{5g^2}{2\pi^2\epsilon}, \qquad
 Z_c=1+\frac{3g^2}{2\pi^2\epsilon},\nonumber\\
 Z_A&=1-\frac{(N_f-30)g^2}{6\pi^2\epsilon}.
 \label{eq:rg_field_gauge_factors}
\end{align}

\begin{figure}
    \centering
    \includegraphics[width=1\linewidth]{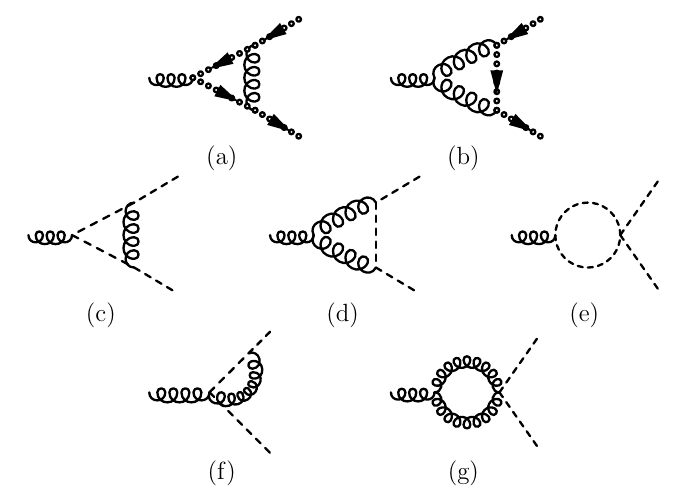}
    \caption{\textbf{One-loop vertex corrections.} (a,b) Corrections to the ghost--gauge vertex. (c--g) Corrections to the Higgs--gauge vertex.}
    \label{fig:feyn_3pt}
\end{figure}

For completeness, we give the Higgs--gauge and ghost--gauge three-point vertex RG factors corresponding to the one-loop diagrams in Fig.~\ref{fig:feyn_3pt}.
These factors are defined through $g_0A_{\mu,0}\partial_\mu Z_0 Z_{0}^\dagger=\mu^{\epsilon/2}Z_{1,Z}g_RA_{\mu,R}\partial_\mu Z_R Z_{R}^\dagger$, and $g_0A_{\mu,0}\partial_\mu \bar c_0c_0=\mu^{\epsilon/2}Z_{1,c}g_RA_{\mu,R}\partial_\mu \bar c_Rc_R$.
The corresponding one-loop vertex RG factors are
\begin{equation}
 Z_{1,Z}=1-\frac{g^2}{2\pi^2\epsilon},\qquad
 Z_{1,c}=1-\frac{3g^2}{2\pi^2\epsilon}.
 \label{eq:rg_vertex_factors}
\end{equation}
They satisfy the one-loop Slavnov--Taylor relation
\begin{equation}
 Z_g=\frac{Z_{1,Z}}{Z_Z Z_A^{1/2}}
 =\frac{Z_{1,c}}{Z_c Z_A^{1/2}}=1+\frac{(N_f-66)g^2}{12\pi^2\epsilon},
 \label{eq:rg_slavnov_taylor}
\end{equation}
providing a consistency check of the field, vertex, and gauge-coupling renormalization.

\begin{figure}
    \centering
    \includegraphics[width=1\linewidth]{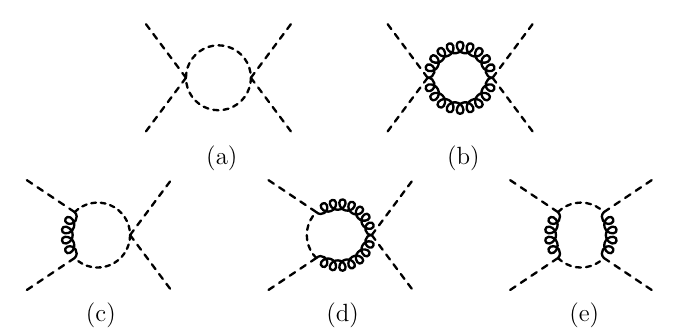}
    \caption{\textbf{One-loop corrections to quartic Higgs interactions.}
    (a) Higgs-loop correction.
    (b) Gauge-loop correction.
    (c--e) Mixed Higgs--gauge corrections.}
    \label{fig:feyn_4pt}
\end{figure}

The one-loop corrections to the quartic Higgs interactions are shown in Fig.~\ref{fig:feyn_4pt}.
For the quartic couplings, we give the equivalent additive counterterms
to avoid ratios of couplings:
\begin{equation}
 \begin{aligned}
 u_{i,0}&=\mu^\epsilon
 \left(u_i+\frac{P_i}{4\pi^2\epsilon}\right),\quad i=1,2,\\
 v_0&=\mu^\epsilon
 \left(v+\frac{P_v}{4\pi^2\epsilon}\right).
 \end{aligned}
 \label{eq:rg_quartic_counterterms}
\end{equation}
The pole polynomials are
\begin{align}
 P_1={}&6g^4-30g^2u_1+4(N_f+1)u_1^2\nonumber\\
 &+2(N_f+4)u_1u_2+2(N_f-1)u_1v\nonumber\\
 &+3u_2^2+2u_2v+v^2,\nonumber\\
 P_2={}&18g^4-30g^2u_2+6u_1u_2\nonumber\\
 &+(N_f+4)u_2^2-2u_2v+(N_f-2)v^2,\nonumber\\
 P_v={}&18g^4-30g^2v+6u_1v\nonumber\\
 &+2(N_f-2)u_2v+4v^2.
 \label{eq:rg_quartic_poles}
\end{align}
Thus $u_i(Z_{u_i}-1)=P_i/(4\pi^2\epsilon)$ and
$v(Z_v-1)=P_v/(4\pi^2\epsilon)$, which specify the corresponding
multiplicative factors.

\begin{figure}
    \centering
    \includegraphics[width=1\linewidth]{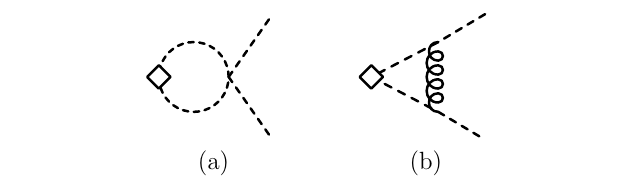}
    \caption{\textbf{One-loop corrections to composite operators.}
    (a) Correction from quartic Higgs interactions.
    (b) Correction from gauge-boson exchange.
    The diamond denotes an insertion of either the charge-$2e$ composite operator or the mass composite operator.}
    \label{fig:feyn_comp}
\end{figure}

For the composite operators
$\hat O_{2e,0}=Z_0\Omega Z_0^T$ and
$\hat O_{m,0}=\operatorname{Tr}(Z_0Z_0^\dagger)$, we define
\begin{equation}
 \hat O_{s,0}=Z_s\hat O_{s,R},\qquad
 Z_s=Z_{\Gamma_s}Z_Z,\quad s=2e,m.
 \label{eq:rg_composite_convention}
\end{equation}
With this convention, $\hat O_{s,R}=Z_{\Gamma_s}^{-1}\hat O_s[Z_R]$.
The one-loop insertion diagrams for these operators are shown in Fig.~\ref{fig:feyn_comp}. 
The corresponding insertion factors are
\begin{align}
 Z_{\Gamma_{2e}}&=1+
 \frac{5g^2-u_1+u_2-4v}{4\pi^2\epsilon},\nonumber\\
 Z_{\Gamma_m}&=1+\frac{1}{4\pi^2\epsilon}
 \bigl[5g^2-(4N_f+1)u_1\nonumber\\
 &\hspace{3em}-(N_f+4)u_2-(N_f-1)v\bigr].
 \label{eq:rg_composite_factors}
\end{align}
Together with $Z_Z$, these factors give the composite-operator scaling
dimensions reported in the Methods.

\section*{Fermionic Parton Pairing}

In this section, we illustrate a possible parton route to the proposed Sp(4) gauge-Higgs description. 
We consider a fractionalization of the electron operator into a bosonic matrix field and a charge-neutral fermion, $c_a = Z_{a\alpha} f_\alpha $, with an implicit sum over the internal index $\alpha$. 
This fractionalization has a local U(4) gauge redundancy because the physical electron operator remains invariant under the transformation
\begin{equation}
    Z \rightarrow Z  \tilde V^\dagger \,,  \quad  f \rightarrow \tilde V f, \quad \tilde V \in  U(4) \,.
\end{equation}
This renders an emergent U(4) gauge structure, which can be written as U(4) $\cong [\rm SU(4) \times U(1)]/ \mathbb{Z}_4$. 

In the charge-$2e$ phase, $Z$ condenses in a locked configuration $Z_{a\alpha}= \rho \delta_{a\alpha}$, which Higgses the U(4) gauge redundancy completely. As a result, the fermionic parton recombines into the physical electron, $c_a = \rho f_a$, with quasiparticle weight given by $|\rho|^2$. 
Hence, the fermionic partons inherit the dispersion of the electron as well as the pairing structure, $\Delta_f = f \Omega f$. 

Outside the Higgs phase, the fermionic parton cannot be mapped directly to the physical electron. 
In this regime, the putative theory is a parton Fermi surface coupled to emergent SU(4) gauge bosons (gluons) and a U(1) gauge boson (photon). 
The unscreened transverse component of gluons (photon) will mediate an attractive (repulsive) interaction in the pairing channel. 
As we show below, when the attractive gluon-mediated interaction dominates, the parton Fermi surface is destabilized toward the pairing $\Delta_f \sim f \Omega f$. 

The exchange process of gauge bosons will effectively generate interaction in the pairing channel. 
Notice that the Fierz identity for the SU(4) generator is 
\begin{eqnarray}
    \sum_i T^i_{\alpha \gamma} T^i_{\beta \delta} = 4 \left( \delta_{\alpha\delta}\delta_{\beta\gamma} - \frac14 \delta_{\alpha\gamma}\delta_{\beta\delta} \right) \,,
\end{eqnarray}
which means that 
\begin{eqnarray}
     \sum_i T^i_{\alpha \gamma'} T^i_{\beta \delta'} P_{\gamma'\delta', \gamma\delta} = -5 P_{\alpha\beta, \gamma \delta} \,,
\end{eqnarray}
where $P_{\alpha\beta, \gamma \delta} = \delta_{\alpha\gamma} \delta_{\beta \delta} - \delta_{\alpha\delta} \delta_{\beta \gamma}$ is the antisymmetric channel.
Including the photon contribution, the gauge bosons exchange process leads to the fermionic parton interaction
\begin{eqnarray}
    \delta V \propto (e^2- 5 g^2) D \int_{k,k'} P_{\alpha\beta, \gamma \delta} f^\dag_\alpha(k) f^{\dag}_\beta(-k) f_\delta(k') f_\gamma(-k'), \nn\\
\end{eqnarray}
where $D>0$ denotes the integral of the unscreened propagator over the exchange momenta. 
In the fractionalized description relevant to the SU(4) Hubbard model, we expect the non-Abelian gauge interaction to dominate over the Abelian one, $5g^2>e^2$. 
In this regime, gauge-boson exchange produces a net attractive interaction in the antisymmetric channel, rendering the parton Fermi surface unstable toward pairing and thereby supporting a finite parton-pairing amplitude, $\Delta_f \sim f \Omega f$, across the transition.  
Related critical Fermi surfaces coupled to non-Abelian gauge fields (e.g., U(2)) have been analyzed in Ref.~\cite{zou2020deconfined}. 
A systematic treatment of a Fermi surface coupled to a U(4) gauge boson, and its interplay with pairing and confinement/Higgs physics, is beyond the scope of the present work and is left for future study.

Crucially, the finite parton pairing plays two roles. 
First, because it is gauge charged, its onset Higgses the emergent U(4) gauge structure down to the subgroup that preserves $\Omega$, namely, Sp(4). 
In this way, the paired parton sector provides a microscopic route by which the low-energy theory reduces to an Sp(4) gauge field coupled to the Higgs field $Z$.
Second, it gaps out the single-electron excitation across the transition.

\begin{figure}
    \centering
    \includegraphics[width=1.0\linewidth]{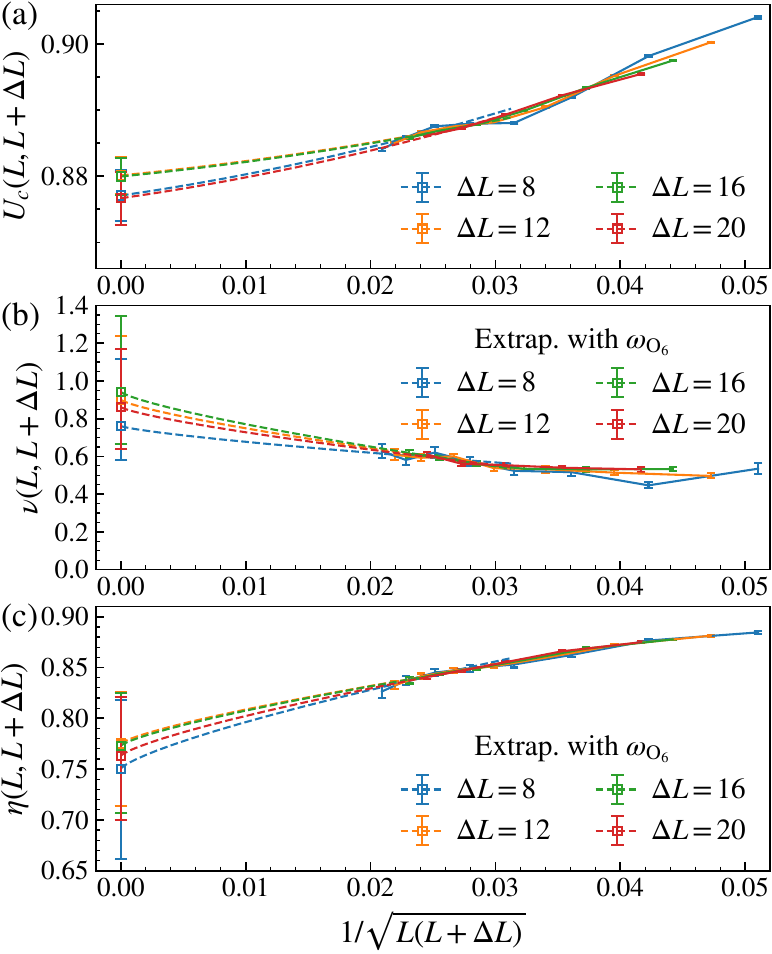}
    \caption{\textbf{Size-pair extrapolation at $1/8$ filling.}
    (a) Crossing estimates $U_c(L,L+\Delta L)$.
    (b,c) Effective exponents $\nu(L,L+\Delta L)$ and $\eta(L,L+\Delta L)$, respectively, for $\Delta L=8,12,16,20$.
    All fits use $L_{\mathrm{eff}}=\sqrt{L(L+\Delta L)}\ge32$.
    Fixed-power scans cover $q\in[1,2]$ for the crossing drift in (a) and $p\in[0.5,2]$ for the exponents in (b,c).
    The leftmost error bars show the scanned intercept envelopes separately for each $\Delta L$.    
    In (a), the points and dashed guides denote envelope midpoints and pointwise midlines.
    In (b,c), the points and dashed curves correspond to $p=\omega_{O(6)}=0.758$ fits; the asymmetric error bars retain the full scanned envelope endpoints.
    The outer range of the four $\Delta L$ intervals in (b,c) gives the envelope in the main text.
}
    \label{fig:extrapolation}
\end{figure}

\section*{Finite-size scaling and extrapolation}

In this section, we detail the finite-size scaling analysis and extrapolation procedures used to extract the critical interaction strength and exponents reported in the main text.

The finite-size scaling analysis is based on the scaling forms of $R_{2e}(U,L)$ and $m_{2e}(U,L)$:
\begin{eqnarray}
    R_{2e}(U,L) &=& f_{R}\!\left((U-U_c)L^{1/\nu}\right) + \mathcal{O}(L^{-\omega})\,,\\
    m_{2e}(U,L) &=& L^{-1-\eta}\,\left[ f_{m}\!\left((U-U_c)L^{1/\nu}\right)+ \mathcal{O}(L^{-\omega})\right]\,, \nonumber
\end{eqnarray}
where $f_R$, $f_m$ are universal scaling functions, and $\mathcal{O}(L^{-\omega})$ terms denote the corrections to scaling.

The data collapse shown in Fig.~3(a) of the main text is performed using a least-squares fitting procedure, in which the scaling function $f_R$ is parameterized as a smooth function of the scaling variable, taken to be a third-order polynomial. 
Due to strong finite-size effects, the collapse is found to be unstable. 
We have also explored alternative polynomial forms and included correction-to-scaling terms; however, the resulting $\chi^2/\mathrm{dof}$ remains significantly larger than unity in all cases.

Instead, we adopt a size-pair crossing analysis to extract the critical interaction strength and effective critical exponents, as described below.
For a pair of system sizes $L_1$ and $L_2$, we select several data points in the vicinity of the crossing and interpolate them to obtain smooth curves $\tilde R_{2e}(U,L_1)$ and $\tilde R_{2e}(U,L_2)$. 
The crossing point of these interpolated curves defines a finite-size estimate of the critical interaction strength, denoted as $U_c(L_1,L_2)$.

From the scaling behavior of $R_{2e}(U,L)$, the slope at criticality satisfies $\partial_U R_{2e}(U_c,L) \propto L^{1/\nu}$. 
This allows us to extract an effective correlation-length exponent for each size pair via 
\eq{
\nu(L_1,L_2)=\frac{\log [L_1/ L_2]}{\log[\partial_U \tilde R_\text{2e}(U^*,L_1)/\partial_U \tilde R_\text{2e}(U^*,L_2)]},
}{}
where $U^*=U_c(L_1,L_2)$.
Similarly, for the anomalous dimension $\eta$, we use the scaling behavior of $m_{2e}$ at criticality, $m_{2e}(U_c,L)\propto L^{-1-\eta}$. 
By interpolation, we obtain smooth curves $\tilde{m}_{2e}(U,L_1)$ and $\tilde{m}_{2e}(U,L_2)$. 
Evaluating these at the crossing point $ U_c(L_1,L_2)$, we extract an effective anomalous dimension for each size pair via
\eq{
\eta(L_1,L_2)=\frac{\log \left[\tilde{m}_{2e}(U^*,L_2)/ \tilde{m}_{2e}(U^*,L_1)\right]}{\log \left[L_1/ L_2\right]}-1\,.
}{}

Specifically, for each size pair, we select five data points in the vicinity of the crossing and perform an interpolation using a third-order polynomial.
Statistical uncertainties of the extracted critical points and exponents are estimated using a bootstrap resampling procedure. As shown in Fig.~\ref{fig:extrapolation}, we consider size pairs of the form $(L, L+\Delta L)$ with $\Delta L = 8, 12, 16$ and $20$.
It is clear that the extracted values have finite-size drift, among these three quantities, $U_c$ and $\eta$ seem more stable than $\nu$, but the trend is similar for different choices of $\Delta L$. 

For each fixed size separation $\Delta L$, we extrapolate
the effective exponents using
\[
X_{\mathrm{eff}}(L_{\mathrm{eff}})
=
X_\infty+A L_{\mathrm{eff}}^{-p},
\qquad X=\nu,\eta,
\]
where $L_{\mathrm{eff}}=\sqrt{L(L+\Delta L)}$.
We scan $p\in[0.5,2]$ to include the leading O(6) correction, $p=\omega_{O(6)}\simeq0.758$, and assess sensitivity to both slower and faster single-power corrections.

Crossing estimates are fitted separately to
\[
U_c(L,L+\Delta L)=U_c+B L_{\mathrm{eff}}^{-q},
\qquad q\in[1,2].
\]
This scan spans inverse-linear to inverse-quadratic drift and includes the O(6) expectation $q=1/\nu_{O(6)}+\omega_{O(6)}\simeq2$.
For $1/8$ filling, we use $\Delta L=8,12,16,20$ and
$L_{\mathrm{eff}}\ge32$.

For each $\Delta L$, we construct an envelope from the fitted means $\pm1\sigma$ over the scanned powers.
Figure~\ref{fig:extrapolation} shows the resulting intercept envelopes separately for each sequence.
The crossing estimates locate the transition near $U_c\simeq0.878$, while the outer ranges of the exponent envelopes give $\nu_\infty\in[0.579,1.342]$ and $\eta_\infty\in[0.661,0.825]$.
The gray bands in Fig.~3 of the main text combine the envelopes from all four $\Delta L$ sequences.
Its representative curves and diamonds instead show the $\Delta L=16$, $p=\omega_{O(6)}=0.758$ fits, which give $\nu_\infty\simeq0.943$ and $\eta_\infty\simeq0.773$.
The quoted ranges characterize extrapolation sensitivity
within this fitting family, rather than statistical
confidence intervals for asymptotic critical exponents.

\section*{Complementary numerical results at $1/8$ filling}
\noindent{\bf Sp(4) symmetry and single-particle gap.}
\begin{figure}
    \centering
    \includegraphics[width=1.0\linewidth]{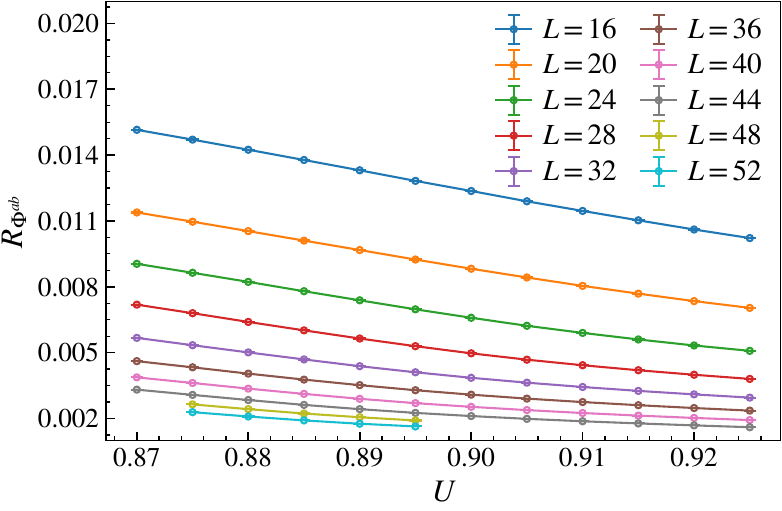}
    \caption{\textbf{Absence of long-range order in the flavor-off-diagonal bilinear channel.} Correlation ratio $R_{\Phi^{ab}}$ for $\Phi^{ab}(i)\equiv c_i^{a\dagger}c_i^{b}$ with $a\neq b$.
    As the system size increases, the correlation ratio decreases systematically toward zero over the entire interaction range. This indicates the absence of long-range order in this channel and therefore no spontaneous symmetry breaking beyond the SU(4) $\rightarrow$ Sp(4) reduction within the charge-$2e$ phase.
    }
    \label{fig:hop_ratio}
\end{figure}
In the charge-$2e$ phase, the pairing state may in principle either preserve or spontaneously break the Sp(4) symmetry. 
To probe this possibility, we introduce the onsite fermion bilinear $\Phi^{ab}(i)\equiv c_i^{a\dagger}c_i^{b}$, which transforms nontrivially under Sp(4) rotations and therefore serves as a diagnostic of possible Sp(4) symmetry breaking. 
We compute the Fourier transform of the exchange kernel, $S_{\Phi^{ab}}(\mathbf q)=\frac{1}{N_s}\sum_{i,j}e^{i\mathbf q\cdot(\mathbf r_i-\mathbf r_j)} \left\langle G_{ij}(\mathbf s)G_{ji}(\mathbf s)\right\rangle_{\mathbf s}$, and define $R_{\Phi^{ab}}=1-S_{\Phi^{ab}}(\mathbf k_{\min})/S_{\Phi^{ab}}(\mathbf0)$, where $\mathbf k_{\min}=(2\pi/L,0)$.
If Sp(4) were spontaneously broken, the correlation function of $\Phi^{ab}(i)$ ($a\neq b$) would develop long-range order, and the associated correlation ratio would approach one in the thermodynamic limit.

As shown in Fig.~\ref{fig:hop_ratio}, however, the correlation ratio decreases toward zero as the system size increases. 
This scaling behavior indicates the absence of long-range order in $\Phi^{ab}(i)$, implying that Sp(4) symmetry remains unbroken. 
We therefore conclude that the Sp(4) symmetry is preserved on both sides of the phase transition.

\begin{figure}
    \centering
    \includegraphics[width=1.0\linewidth]{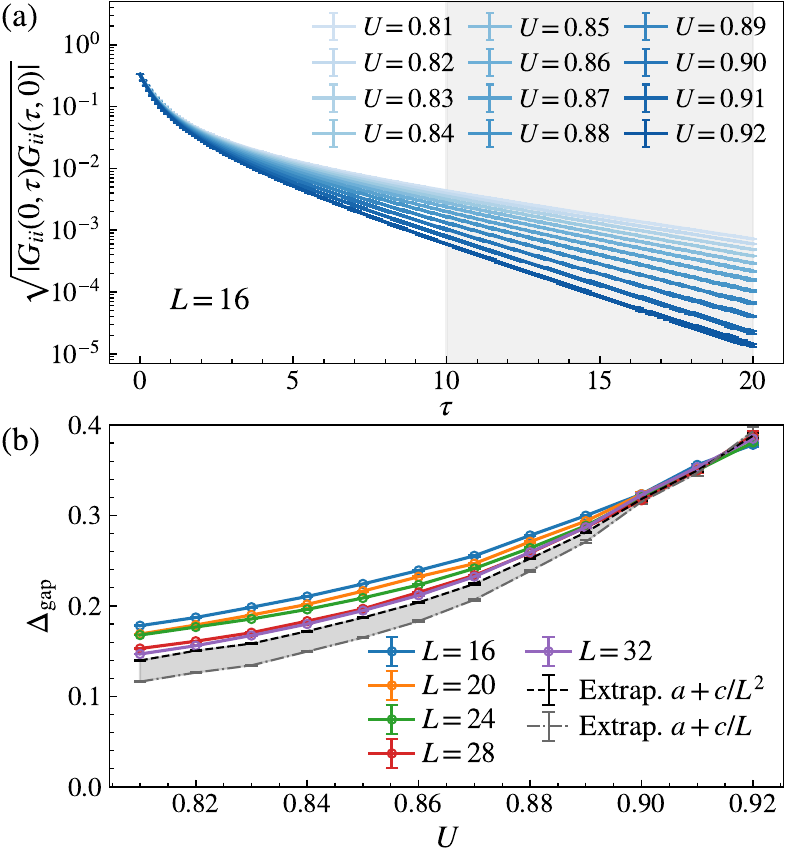}
    \caption{\textbf{Single-particle gap across the phase transition.} (a) Imaginary-time Green’s function for different interaction strengths at $L=16$. The data exhibit clear exponential decay, from which the single-particle gap is extracted by fitting the long-time regime $\tau>10$ (shaded region). 
    (b) Extracted single-particle gap as a function of interaction strength for different system sizes. To extrapolate to the thermodynamic limit, we employ two fitting forms, $a+c/L$ and $a+c/L^2$ (dotted lines). Both extrapolations yield a finite single-particle gap on either side of the phase transition, indicating that the single-particle excitation remains gapped throughout the transition.
    Here, we set $2\Theta = 70 + L$ and evaluate the time-displaced Green’s function in the central region. }
    \label{fig:gap_function}
\end{figure}

Next, we investigate the single-particle gap across the phase transition. Specifically, we compute the time-displaced Green's functions $G_{ij}(\tau,0) \equiv \langle c_i(\tau)c_j^\dagger(0)\rangle$ and $G_{ij}(0,\tau) \equiv -\langle c_j^\dagger(\tau)c_i(0)\rangle$. 
Since the simulation is performed in the canonical ensemble, the particle number is fixed and the chemical potential is not directly accessible. 
We therefore consider the combination $G_{\rm{eff}}(\tau)=\sqrt{|G_{ii}(\tau, 0)G_{ii}(0,\tau)|}$ for which the chemical-potential dependence is eliminated. 
At large $\tau$, this quantity should behave as $G_{\rm eff}(\tau)\propto e^{-\Delta_{\rm gap}\tau}$, where $\Delta_\text{gap}$ is the single-particle gap.

As shown in Fig.~\ref{fig:gap_function}(a), $G_{\rm eff}$ exhibits clear exponential decay at large $\tau$. 
By fitting this long-time behavior, we extract the single-particle gap $\Delta_\text{gap}$. 
As shown in Fig.~\ref{fig:gap_function}(b), the extrapolated single-particle gap remains finite in the thermodynamic limit on both sides of the phase transition, confirming that single-particle excitations remain gapped throughout the transition.

\begin{figure}[t]
    \centering
    \includegraphics[width=0.95\linewidth]{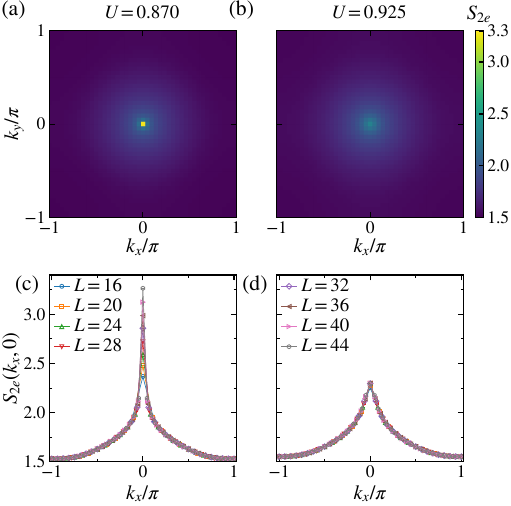}
    \caption{\textbf{Momentum-resolved charge-$2e$ pairing correlations at $1/8$ filling.}
    The structure factor $S_{2e}(\mathbf{k})$ follows the operator definition and normalization of Eq.~2 of the main text.
    (a,b) Full Brillouin-zone maps at $L=44$ for $U=0.870$ and $0.925$, respectively, using a shared linear color scale.
    (c,d) Corresponding cuts along $k_y=0$, from $k_x=-\pi$ to $+\pi$, for $L=16,20,24,28,32,36,40,44$.
    Each size has the same color and marker in both panels; the shared size legend is split between them.
    Markers show native discrete momenta, and lines join adjacent measurements without smoothing.
    The periodic endpoints represent the same momentum.
    Error bars denote one standard error from an independent-chain jackknife of the bridge-reweighted ratio, using the final half of complete, aligned measurement bins; some are smaller than the markers.
    Within the sizes and two interactions examined, the dominant maximum remains at $\mathbf{k}=0$, with no competing finite-momentum peak resolved.
}
    \label{fig:pairing_momentum}
\end{figure}

\begin{figure}[t]
    \centering
    \includegraphics[width=0.95\linewidth]{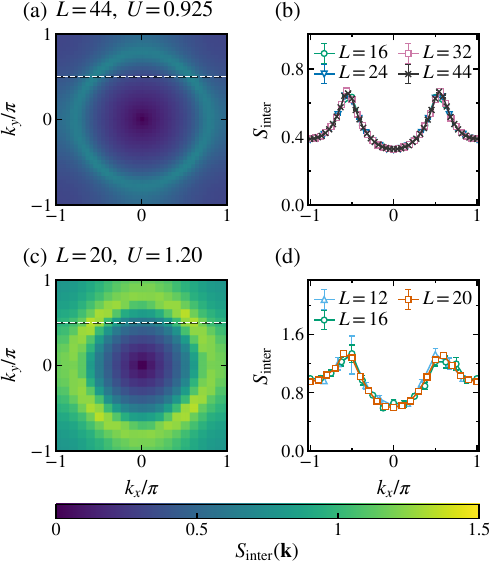}
    \caption{\textbf{Connected interflavor density structure factor at $1/8$ filling }. 
    The definition and normalization follow Eq.~\eqref{eq:si_interflavor_definition}.
    (a,b) Results at $U=0.925$: an $L=44$ momentum map and cuts for $L=16,24,32,44$.
    (c,d) Results at $U=1.20$: an $L=20$ momentum map and cuts for $L=12,16,20$.
    White dashed lines indicate $k_y=\pi/2$; both momentum maps share the same color scale.
    Error bars denote independent-chain jackknife standard errors.
    Possible bias from incomplete mixing, particularly for $U=1.20$, is not captured by these errors.
}
    \label{fig:density_momentum}
\end{figure}

\noindent{\textbf{Momentum-resolved pairing and density correlations at $1/8$ filling}.}
\label{sec:momentum_eighth}
To further characterize the charge-$4e$ superconducting regime, we examine the momentum dependence of the charge-$2e$ pairing and density correlations.
The correlation ratio $R_{2e}$ used in the main text probes the uniform pairing channel. 
We therefore also inspect the pairing structure factor $S_{2e}(\mathbf{k})$ throughout the Brillouin zone to check for possible finite-momentum pairing.

Figure~\ref{fig:pairing_momentum} shows representative results at $U=0.870$ and $0.925$, on opposite sides of the transition region.
For $L=44$, the dominant peak is located at $\mathbf{k}=\mathbf{0}$ at both couplings, with no additional peaks resolved at nonzero wavevectors.
Cuts along $k_y=0$ reveal contrasting size dependence: the zero-momentum peak grows with $L$ at $U=0.870$, whereas its size dependence is much weaker at $U=0.925$. 
Together with the finite-size behavior of $R_{2e}$, these observations support the suppression of charge-$2e$ long-range order on the stronger-coupling side and provide no indication of finite-momentum pairing within the sizes and couplings examined.

We study the density channel using the connected interflavor
correlation function and its structure factor,
\begin{align}
C_{\mathrm{inter}}(\mathbf r)
&=\frac{1}{N_s}\sum_i\sum_{a\ne b}
\left[
\left\langle n_{ia}n_{i+\mathbf r,b}\right\rangle-f^2
\right], \nonumber\\
S_{\mathrm{inter}}(\mathbf k)
&=\sum_{\mathbf r}e^{i\mathbf k\cdot\mathbf r}
C_{\mathrm{inter}}(\mathbf r).
\label{eq:si_interflavor_definition}
\end{align}
where $n_{ia}=c_{ai}^{\dagger}c_{ai}$, $N_s=L^2$, and $f=N_e/(4N_s)$ is the filling per flavor.
The sum over $a\ne b$ includes the twelve ordered distinct-flavor pairs. 
The uniform background $f^2$ is subtracted for each pair, and no additional normalization is applied in the Fourier transform.

Figure~\ref{fig:density_momentum}(a,b) shows the results at $f=1/8$ and $U=0.925$.
The $L=44$ momentum map displays broad maxima without resolved sharp ordering peaks. 
Cuts along $k_y=\pi/2$ for $L=16,24,32,44$ show little systematic size dependence.
Figure~\ref{fig:density_momentum}(c,d) extends the density-channel check to $U=1.20$, with an $L=20$ momentum map and cuts for $L=12,16,20$. 
These data likewise show broad maxima rather than resolved sharp
ordering peaks. 
However, sampling becomes more challenging at larger $U$ because of slower mixing.

Near $U=0.925$, these density measurements, together with the pairing
results in Fig.~\ref{fig:pairing_momentum} and the finite-size behavior of the long-distance
quartet correlations in the main text, support a charge-$4e$
superconducting regime without detectable charge-$2e$ or density-wave order in the measured channels and parameter window.

\begin{figure}[t]
    \centering
    \includegraphics[width=0.95\linewidth]{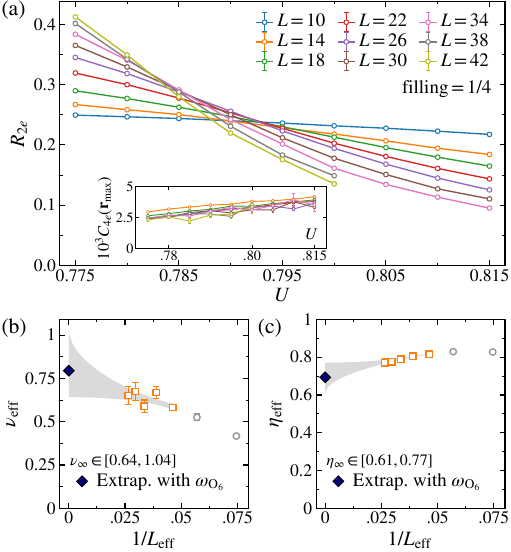}
    \caption{\textbf{Pairing correlations and effective critical exponents at $1/4$ filling.}
(a) Pair correlation ratio $R_{2e}$ versus $U$ for $L=10,14,\ldots,42$.
Inset: largest-separation quartet correlation
$10^3 C_{4e}(\mathbf r_{\max})$, with
$\mathbf r_{\max}=(L/2,L/2)$, for $L=14,18,\ldots,42$.
(b,c) Effective exponents $\nu_{\mathrm{eff}}$ and $\eta_{\mathrm{eff}}$
obtained from crossings of size pairs $(L,L+8)$, plotted against
$1/L_{\mathrm{eff}}$, where $L_{\mathrm{eff}}=\sqrt{L(L+8)}$.
Orange squares enter the extrapolation ($L_{\mathrm{eff}}>20$);
gray circles denote the excluded pairs.
Gray bands show the outer envelopes of fitted means $\pm1\sigma$ for $X_{\mathrm{eff}}=X_\infty+A L_{\mathrm{eff}}^{-p}$, scanning fixed correction powers $p\in[0.5,2]$.
Diamonds at $1/L_{\mathrm{eff}}=0$ indicate representative intercepts obtained with $p=\omega_{\mathrm{O}(6)}=0.758$: $\nu_\infty=0.795$ and $\eta_\infty=0.694$.
The gray bands retain the full scanned uncertainty envelopes.
These envelopes characterize correction-power sensitivity rather than confidence intervals from a single fit.
}
    \label{fig:quarter_pair_quartet}
\end{figure}

\begin{figure}
    \centering
    \includegraphics[width=1.0\linewidth]{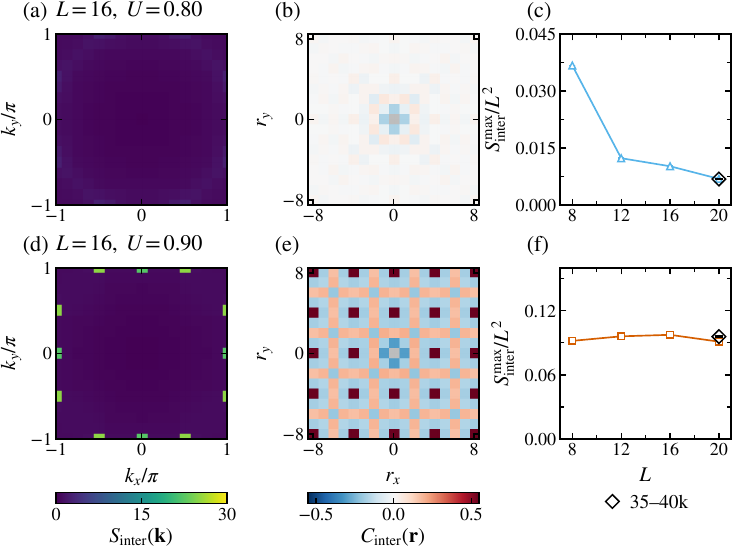}
    \caption{\textbf{$C_4$-averaged connected interflavor density correlations at $1/4$ filling.} Data are shown for $U=0.80$ (top row) and $U=0.90$ (bottom row). 
    (a,d) Momentum-space structure factors and (b,e) real-space correlations at $L=16$.
    The maps share a color scale within each column; the onsite values are masked in the real-space panels.
    (c,f) Normalized maxima $S_{\mathrm{inter}}^{\max}(L)/L^2$, defined in Eq.~\eqref{eq:si_interflavor_peak}, versus $L$.
    Colored points use the retained-half sampling windows; black diamonds additionally show results from the late $35$--$40$k-sweep window at $L=20$.
    Error bars are jackknife estimates over 48 independent chains, with $C_4$ averaging performed within each chain. Pronounced CDW signatures at $U=0.90$ coexist with slow orientation mixing and residual late-time drift.
}
    \label{fig:quarter_density}
\end{figure}

\section*{Additional numerical results at $1/4$ filling}
\label{sec:quarter_filling}

To examine the dependence on filling and provide context for our focus on $1/8$ filling, we also study the $\tfrac{1}{4}$-filled model.
As shown in Fig.~\ref{fig:quarter_pair_quartet}(a), the charge-$2e$ correlation ratio exhibits crossings near $U\simeq0.79$, while the long-distance quartet correlations remain substantial across the crossing region. 
These observations indicate a suppression of uniform charge-$2e$ coherence in a regime with persistent quartet correlations.

Figures~\ref{fig:quarter_pair_quartet}(b,c) show the corresponding effective exponents at quarter filling.
Using $\Delta L=8$, $L_{\mathrm{eff}}>20$, and the same fixed-power scan $p\in[0.5,2]$, we obtain $\nu_\infty\in[0.642,1.039]$ and $\eta_\infty\in[0.606,0.771]$.
The representative $p=\omega_{O(6)}=0.758$ fits give $\nu_\infty\simeq0.795$ and $\eta_\infty\simeq0.694$.
Thus, a large effective pairing anomalous dimension is also observed at quarter filling, although the different size windows and remaining drift do not establish identical asymptotic exponents at the two fillings.

At larger interactions, the quarter-filled data show enhanced density-wave correlations. We use the connected interflavor correlations defined in Eq.~\eqref{eq:si_interflavor_definition}, with $f=1/4$.
For Fig.~\ref{fig:quarter_density}, we average the real- and momentum-space correlations over the four lattice
rotations within each independent chain before jackknife
analysis. Below, $C_{\mathrm{inter}}$ and $S_{\mathrm{inter}}$
denote these $C_4$-averaged quantities.
The normalized peak height plotted in panels (c,f) is
\begin{equation}
\frac{S_{\mathrm{inter}}^{\max}(L)}{L^2}
=
\frac{1}{L^2}
\max_{\mathbf{k}}
S_{\mathrm{inter}}(\mathbf{k};L),
\label{eq:si_interflavor_peak}
\end{equation}
where the maximum is taken over the discrete lattice momenta.

At $U=0.90$, the $L=16$ structure factor develops pronounced peaks at $\mathbf{Q}_x=(\pi/2,\pi)$, $\mathbf{Q}_y=(\pi,\pi/2)$, and symmetry-related momenta [Fig.~\ref{fig:quarter_density}(d)], accompanied by spatially modulated real-space correlations [Fig.~\ref{fig:quarter_density}(e)]. 
The normalized maximum decreases with increasing size at $U=0.80$ [Fig.~\ref{fig:quarter_density}(c)], whereas it remains substantial with much weaker size dependence over $L=8,12,16$ at $U=0.90$ [Fig.~\ref{fig:quarter_density}(f)].
For $L=20$, the peak height estimated from sweeps $35{,}000$--$40{,}000$ differs from that obtained using the second half of the run, indicating residual sampling drift.

The pronounced density correlations at quarter filling make its phase diagram more involved.
At $1/8$ filling, the momentum-resolved measurements in Fig.~\ref{fig:density_momentum} show no comparable sharp ordering peaks over the sizes and couplings examined, subject to the stronger-coupling sampling limitations discussed above.
The smaller particle number at this filling also reduces the computational cost and permits larger lattices.
These considerations motivate our focus on $1/8$ filling for
the systematic study of the charge-$2e$--charge-$4e$ transition.

\clearpage

\end{document}